\documentclass[11pt,twoside,a4paper,cmspaper,final,collab]{cms-tdr}
\def\svnVersion{788ad99-D}\def\svnDate{2026/05/01}\def\cmsCernNoTag{CERN-EP-2026-010}\def\cmsCernDate{\today}\def\cmsMessage{Submitted to Physical Review D}
\begin{document}\cmsNoteHeader{HIG-20-012}

\providecommand{\PY}{\HepParticle{Y}{}{}\xspace}
\providecommand{\PA}{\HepParticle{A}{}{}\xspace}
\newcommand{\HH}{\ensuremath{\PH\PH}\xspace} 
\newcommand{\bbbb}{\ensuremath{\PQb\PAQb\PQb\PAQb}\xspace}
\newcommand{\mX}{\ensuremath{m_{\PX}}\xspace}
\newcommand{\mY}{\ensuremath{m_{\PY}}\xspace}
\newcommand{\Hmssm}{\ensuremath{\PH_{\mathrm{MSSM}}}\xspace}
\newcommand{\Amssm}{\ensuremath{\PA_{\mathrm{MSSM}}}\xspace}
\newcommand{\Hs}{\ensuremath{\PH_{\text{singlet}}}\xspace}
\newcommand{\As}{\ensuremath{\PA_{\text{singlet}}}\xspace}
\newcommand{\mH}{\ensuremath{m_{\PH}}\xspace}
\newcommand{\Xreco}{\ensuremath{\PX_{\text{reco}}}\xspace}
\newcommand{\Yreco}{\ensuremath{\PY_{\text{reco}}}\xspace}
\newcommand{\Hreco}{\ensuremath{\PH_{\text{reco}}}\xspace}
\newcommand{\mXreco}{\ensuremath{m_{\text{Xreco}}}\xspace}
\newcommand{\mYreco}{\ensuremath{m_{\text{Yreco}}}\xspace}
\newcommand{\mHreco}{\ensuremath{m_{\text{Hreco}}}\xspace}
\newcommand{\XHY}{\ensuremath{\PX \to \PY \PH \to \bbbb}\xspace}
\newcommand{\Xhy}{\ensuremath{\PX \to \PY \PH}\xspace}
\newcommand{\XHH}{\ensuremath{\PX \to \PH \PH \to \bbbb}\xspace}
\newcommand{\Xhh}{\ensuremath{\PX \to \PH \PH}\xspace}
\newcommand{\Hbb}{\ensuremath{\PH \to \PQb\PAQb}\xspace}
\newlength\cmsTabSkip\setlength{\cmsTabSkip}{1ex}
\newlength\cmsFigWidth
\ifthenelse{\boolean{cms@external}}
{
    \providecommand{\cmsLeft}{upper\xspace}
    \providecommand{\cmsRight}{lower\xspace}
    \setlength{\cmsFigWidth}{\columnwidth}
}
{
    \providecommand{\cmsLeft}{left\xspace}
    \providecommand{\cmsRight}{right\xspace}
    \setlength{\cmsFigWidth}{0.66\textwidth}
}

\cmsNoteHeader{HIG-20-012}

\title{Search for a new heavy scalar resonance decaying into the Higgs boson and a new scalar particle in the \texorpdfstring{\bbbb}{b bbar b bbar} final state using proton-proton collisions at \texorpdfstring{$\sqrt{s}=13\TeV$}{sqrt(s) = 13 TeV}}

\date{\today}

\abstract{
A search for a new heavy scalar resonance (\PX) decaying into the 125\GeV standard model Higgs boson (H) and a new scalar particle (\PY) in proton-proton collisions at a center-of-mass energy of 13\TeV is presented. The analysis is performed using a data sample corresponding to an integrated luminosity of 138\fbinv collected with the CMS detector during LHC Run~2. The \bbbb final state is used as a probe to search for phenomena beyond the standard model where, in the $\PX \to \PY\PH$ process, the Y and H each decay into a bottom quark-antiquark pair. A range of masses from 400\GeV to 1.6\TeV for the resonance \PX and from 60\GeV to 1.4\TeV for the scalar \PY is investigated. The observations are in agreement with the background-only hypothesis. The largest excess, with a local (global) significance of 3.47\,(2.44) standard deviations, is observed for hypothetical \PX and \PY masses of 600 and 400\GeV, respectively. Upper limits at 95\% confidence level on the production cross section times branching fraction are presented for signal mass hypotheses in the range of the search. Results are interpreted within the next-to-minimal supersymmetric standard model scenario.
}

\hypersetup{%
pdfauthor={CMS Collaboration},%
pdftitle={Search for a new heavy scalar resonance decaying into the Higgs boson and a new scalar particle in the bbbb final state using proton-proton collisions at sqrt(s) = 13 TeV},%
pdfsubject={CMS},%
pdfkeywords={CMS, HH, bbbb, NMSSM, Higgs}%
}%

\maketitle

\section{Introduction}
The discovery of the Higgs boson (\PH) by the ATLAS and CMS Collaborations at the CERN LHC~\cite{ATLASdiscovery,CMSdiscovery,Chatrchyan:2013lba} was the confirmation of a cornerstone of the standard model (SM) of particle physics. At the same time, the Higgs boson could be a portal for the discovery of new physics.
Many proposed models of physics beyond the SM (BSM) extend the SM scalar sector with additional fields, resulting in a rich phenomenology with more scalar particles. Examples of such models are the next-to-minimal supersymmetric SM (NMSSM)  \cite{Ellwanger:2009dp,Maniatis:2009re}, the two-real-scalar-singlet extension of the SM \cite{Robens:2019kga}, and models with warped extra dimensions~\cite{Carvalho:2014lsg}.

The analysis described here explores one possible phenomenological scenario of these theoretical models: the production of a BSM resonance that decays into the SM Higgs boson and another scalar, \Xhy, where Y may be either a BSM scalar or the SM Higgs boson. This is motivated in, for example, NMSSM \cite{Ellwanger:2017skc}, where a scalar \Hmssm or pseudoscalar \Amssm can decay into the SM Higgs boson and a scalar \Hs or pseudoscalar \As, respectively. The branching fraction for this process $B(\Hmssm \to \PH + \Hs)$ or $B(\Amssm \to \PH + \As)$ can be as high as 50\% \cite{Kang:2013rj,King:2014xwa,Carena:2015moc,Ellwanger:2015uaz,Costa:2015llh}.  Previous searches for this process have been published by the ATLAS Collaboration, using the $\bbbar \PGg\PGg$ final state~\cite{ATLAS:2024auw}, and by the CMS Collaboration, using the $\bbbar \PGg\PGg$~\cite{CMS:2023boe}, $\bbbar \PGt\PGt$~\cite{CMS:2021yci}, \bbbb with Lorentz-boosted topology~\cite{CMS:2022suh} final states, and a combination of these three channels~\cite{CMS:2024phk}.

The decay \Xhh is of particular interest, since it can be the manifestation of Randall--Sundrum radion or a massive Kaluza--Klein graviton production in models with warped extra dimensions \cite{Csaki:2007ns, salam}, as well as the charge-parity-even scalar predicted by NMSSM~\cite{Barbieri:2013hxa}. Multiple searches for Higgs boson pair production have been published by the ATLAS and CMS Collaborations in the $\bbbb$~\cite{hh4b,Aaboud:2018knk,ATLAS:HH4b}, $\bbbar\PGt\PGt$~\cite{bbtautau,Aaboud:2018bun}, $\bbbar\PGg\PGg$~\cite{bbgammagamma}, $\bbbar\ell\nu\ell\nu$~\cite{bbWW,Aaboud:2018bun}, and $\bbbar ZZ$ \cite{CMS:2020jeo} channels, in WWWW, WW$\PGt\PGt$, and $\PGt\PGt\PGt\PGt$ final states~\cite{CMS:2022kdx}, and in combinations of multiple channels~\cite{ATLAS:2023vdy, ATLAS:2024ish, CMS:2024phk}.

\begin{figure}[hbt!]
\centering
{\includegraphics[width=\cmsFigWidth] {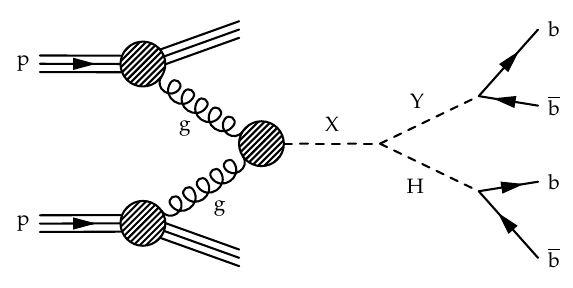}}
\caption{\label{fig:feyn}
    Depiction of the process under investigation, $\XHY$.
}
\end{figure}

In order to be sensitive to a broad range of scenarios, a model-independent approach is followed in this analysis.
Figure~\ref{fig:feyn} shows an example of X production, in which a heavy scalar resonance X is produced via gluon fusion and then decays into an H and another scalar, Y, with both of them decaying into a $\bbbar$ pair. This analysis makes use of the Higgs boson decay mode with the largest branching fraction, \Hbb, providing strong motivation for this channel.

This paper presents a search for X resonances with masses from 400\GeV up to 1.6\TeV and a scalar Y with masses between 60\GeV and 1.4\TeV.
The boundaries of the search region are defined to encompass the region of phase space where this analysis provides greater sensitivity than other channels probing similar resonances; lower masses are covered by the $\bbbar\PGg\PGg$ channel, while higher masses are targeted by analyses using Lorentz-boosted, merged $\PQb$~jet topologies.
The signal is extracted from a two-dimensional fit using the masses of the reconstructed X and Y scalars, \mXreco and \mYreco, respectively.
The data samples used for the search correspond to an integrated luminosity of 138\fbinv collected by the CMS experiment in the years 2016--2018.
An approach based on control samples in data, utilizing a boosted decision tree (BDT) for reweighting kinematic variables, is implemented for the estimation of the background.
A simultaneous fit to the data samples collected from each year is used to infer the results.
The tabulated results of the analysis are provided in the HEPData record~\cite{hepdata}.

\section{The CMS detector}

The CMS apparatus~\cite{CMS:2008xjf,CMS:2023gfb} is a multipurpose, nearly hermetic detector, designed to trigger on~\cite{CMS:2020cmk,CMS:2016ngn} and identify electrons, muons, photons, and (charged and neutral) hadrons~\cite{CMS:2020uim,CMS:2018rym,CMS:2014pgm}. A global ``particle-flow'' (PF) algorithm~\cite{CMS:2017yfk} aims to reconstruct all individual particles in an event (PF candidate), combining information provided by the all-silicon inner tracker and by the crystal electromagnetic calorimeter (ECAL), and a brass and scintillator hadron calorimeter (HCAL), operating inside a 3.8\unit{T} superconducting solenoid, with data from the gas-ionization muon detectors embedded in the flux-return yoke outside the solenoid. The PF candidates are used to identify and reconstruct \PGt leptons, jets, and missing transverse momentum~\cite{CMS:2018jrd,CMS:2016lmd,CMS:2019ctu}.

The HCAL cells have widths of 0.087 in pseudorapidity ($\eta$) and 0.087 radians in azimuthal angle ($\phi$), in the region $\abs{\eta} < 1.74$. In the $\eta$-$\phi$ plane, and for $\abs{\eta} < 1.48$, the HCAL cells map on to $5{\times}5$ arrays of ECAL crystals to form calorimeter towers that project radially outwards from a location close to the nominal interaction point.
For $\abs{\eta} > 1.74$, the transverse dimensions of these towers increase progressively, up to a maximum of 0.174 in both $\Delta \eta$ and $\Delta \phi$.

Events of interest are selected using a two-tiered trigger system. The first level (L1), composed of custom hardware processors, uses information from the calorimeters and muon detectors to select events at a rate of around 100\unit{kHz}, within a fixed latency of 4\mus~\cite{CMS:2020cmk}. The second level, known as the high-level trigger (HLT), consists of a farm of processors running a version of the full event reconstruction software optimized for fast processing, and reduces the event rate to around 1\unit{kHz} before data storage~\cite{CMS:2016ngn,CMS:2024aqx}.

\section{Event reconstruction}

In the PF event reconstruction, the identification of the particle type (photon, electron, muon, charged or neutral hadron) plays an important role in the determination of the particle direction and energy.
For each event, hadronic jets are clustered from PF candidates using the infrared and collinear-safe anti-\kt algorithm~\cite{Cacciari:2008gp, Cacciari:2011ma} with a distance parameter of 0.4. Jet momentum is determined as the vectorial sum of all particle momenta in the jet, and is found from simulation to be, on average, within 5 to 10\% of the true momentum over the entire transverse momentum (\pt) spectrum and detector acceptance.
The jets in the events are required to have $\abs{\eta} < 2.4$ for data-taking year 2016 and $\abs{\eta} < 2.5$ for 2017--2018. The $\abs{\eta}$ range for jet reconstruction was increased as a result of the installation of the new pixel detector at the beginning of 2017~\cite{Vormwald:2019pyc}.

Additional proton-proton (pp) interactions within the same or nearby bunch crossings (pileup) can contribute additional tracks and calorimetric energy depositions to the jet momentum.
To mitigate this effect, charged particles identified to be originating from pileup vertices are discarded and an offset correction is applied to account for remaining contributions~\cite{CMS:2020ebo}.
Jet energy corrections are derived from simulation to bring the measured response of jets to that of particle level jets on average. In situ measurements of the momentum balance in dijet, photon+jet, Z+jet, and multijet events are used to determine any residual differences between the jet energy scale in data and in simulation, and appropriate corrections are made~\cite{CMS:2016lmd}.
Additional selection criteria are applied to each jet to remove jets potentially dominated by anomalous contributions from various subdetector components or reconstruction failures~\cite{CMS:2020ebo}.

In order to suppress the background originating from leptonic decays of the top quark, events are vetoed if they contain an isolated electron (muon) candidate with $\pt> 15\;(10)\GeV$ and $\abs{\eta}<2.4$. Criteria for identifying these objects include passing a multivariate identification discriminant and passing selections on isolation and impact parameter with respect to the primary interaction vertex observables~\cite{CMS:2020uim,CMS:2018rym}.

Jets originating from a bottom quark (\PQb~jets) are identified using the \textsc{DeepJet} algorithm~\cite{Bols:2020bkb} and the loose or medium \PQb~tagging working point~\cite{BTV-16-002}.
The working points are defined by the rate of jets that are initiated from light flavor (up, down, or strange) quarks or gluons, but are incorrectly identified as \PQb~jets.
The loose and medium working points are defined such that the misidentification rates of light flavor and gluon (charm) jets are about 10 and 1 (40 and 10)\%, respectively.
The loose (medium) working point has a \PQb~tagging efficiency in the range 85-90\,(68-78)\%, depending on the data-taking year.

\section{Simulated samples}
\label{sec:sims}

Signal samples have been generated at leading-order (LO) in perturbative quantum chromodynamics (QCD) using \MGvATNLO~\cite{Alwall:2014hca} 2.4.2 (2.6.0) for 2016 (2017--2018).
Signal events are generated with the width of the X and Y scalar resonances set to 1\MeV, which is negligible compared to the experimental resolution.
The parton distribution function (PDF) set NNPDF3.0 with next-to-LO (NLO) accuracy~\cite{NNPDF:2014otw} and NNPDF3.1 with next-to-NLO accuracy~\cite{NNPDF:2017mvq} are used in the signal generation in 2016 and 2017--2018 samples, respectively.

Background samples are generated to qualitatively assess the processes that contribute to the background composition and to optimize the event selection. However, the analysis uses control samples in data to model the background, described in Section~\ref{sec:bkgModeling}.
Simulated events from QCD multijet processes are generated with \MADGRAPH at LO precision, with up to four additional partons.
The production of pairs of top quarks is simulated at NLO precision with \POWHEG~2.0~\cite{Alioli:2008gx, Nason:2004rx}, and events are normalized to the next-to-NLO precision theoretical cross section~\cite{Czakon:2011xx}.
Events with a \PZ boson pair decaying into hadrons, with up to one additional jet emitted at the matrix element level, are simulated with \MADGRAPH at NLO precision and normalized to the cross section at NLO precision~\cite{Campbell:2011bn}.

For all simulated samples, the parton shower and hadronization is performed with \PYTHIA version 8.226 (2016) and 8.230 (2017--2018)~\cite{Sjostrand:2014zea}, and \GEANTfour~\cite{GEANT4:2002zbu} is used to model the CMS detector response.

Contributions due to pileup are simulated and added to the signal and background samples.
The simulated events, including both the signal and background samples, are weighted to match the number of pileup observed in the data.

\section{Event selection and categorization}
Events passing a combination of L1 triggers that require the presence of one or more jets are of interest to this analysis; between 5 and 18 L1 trigger algorithms are used in each individual CMS data-taking period. A range of jet \pt and \HT (the scalar sum of the \pt of all jets in the event) thresholds are used to select events.
The L1 triggers requiring at least one jet in the event have a \pt threshold on a jet ranging from 170 to 200\GeV.
Triggers based on \HT selection use values for the threshold from 280 to 500\GeV.
The L1 triggers requiring four jets have \HT thresholds ranging from 250 to 340\GeV with individual \pt selection for each jet; the minimum \pt requirement can be as low as 35\GeV for the lowest \pt jet.
Other L1 trigger algorithms are also utilized, including those with requirements for at least two or at least three jets.

Dedicated HLT algorithms targeting the signal process have been used to refine the trigger-level selection. The HLT criteria include selections on the \pt and the \HT of the \PQb~tagged jets in the event.
Each HLT algorithm requires the presence of at least three \PQb~tagged jets.
The Combined Secondary Vertex (\textsc{CSV}) algorithm~\cite{BTV-16-002} (2016--2017) and the \textsc{DeepCSV} algorithm~\cite{Bols:2020bkb} (2018) are used for \PQb~tagging at the HLT level.
The thresholds of the \pt and \HT
selection criteria were increased in 2017--2018, with respect to those in 2016, to sustain data-taking in higher pileup conditions.
An offline selection at 90\% of the HLT turn-on curves, close to the plateau where the online selection reaches 100\% efficiency, for the \pt and \HT of \PQb~tagged jets is made. These offline selection criteria are listed in Table~\ref{tab:HLTcuts}.

\begin{table}[!htb]
\topcaption{\label{tab:HLTcuts} Offline selection thresholds at 90\% of HLT trigger turn-on curves. The values represent the lower bounds on \pt and \HT for the four highest \pt jets in an event for each data-taking year. The values in the table are in units of {\GeVns}. For data-taking year 2016, events that pass either of two trigger algorithms are used.}
\centering
\begin{scotch}{l r r r r}
    & 2016 (1) & 2016 (2) & 2017 & 2018 \\
	\hline
    $p_{\mathrm{T,1}}$  & 110          & 65          & 95   & 95   \\
    $p_{\mathrm{T,2}}$  & 110          & 65          & 80   & 80   \\
    $p_{\mathrm{T,3}}$  & 45          & 65          & 60   & 60   \\
    $p_{\mathrm{T,4}}$  & 45          & 65          & 60   & 60   \\
    \HT               & 320         & 320         & 420  & 450
\end{scotch}
\end{table}

Further offline selection criteria are implemented in order to target the signal process.
To reduce the number of physics objects that are poorly reconstructed, jets with $\pt > 50\GeV$ are required to pass the tight working point of the PF jet identification algorithm~\cite{CMS:2017wyc}, which has an efficiency of 99\%. For jets with \pt between 30 and 50\GeV, a separate requirement is applied, specifically that the jet must pass the medium working point of the pileup discriminant~\cite{CMS:2020ebo}, which has an efficiency of about 90\%.
To take into account undetected energy carried by a neutrino originating from \PQb hadron decays, the jet \pt is corrected with a multivariate regression technique, as described in Ref.~\cite{CMS:2019uxx}.

The four jets with the highest \textsc{DeepJet} \PQb~tagging score are used to identify the X, Y, and H candidates. The four jets are required to be the same objects that passed the HLT requirements. This selection requires the angular separation, $\Delta R = \sqrt{\smash[b]{(\Delta\eta)^2+(\Delta\phi)^2}}$, between the triggered object and the reconstructed jet to be less than 0.5.
Two datasets are created with different \PQb~tagging score requirements.
The main dataset (4\PQb) has the potential to contain the signal processes, and is formed by requiring four jets that satisfy the medium working point of the \textsc{DeepJet} discriminant. The background-dominated dataset (3\PQb) is used for modeling the background, and is formed by requiring that three jets satisfy the medium working point and that one jet passes the loose, but fails the medium, working point.

The H and Y candidates are reconstructed by pairing the four jets in the 3\PQb and 4\PQb samples. Out of the possible jet pairs, the one with an invariant mass closest to 125\GeV is selected as the \PH candidate and the remaining pair is the Y candidate, with mass \mYreco. As a consequence of this choice all events used in this analysis will meet the condition $\abs{\mYreco-125\GeV} > \abs{\mHreco -125\GeV}$. The efficiency of correctly paired candidates is determined using the simulated samples.
The angular separation $\Delta R$ between the generated Higgs boson (generated Y) and the reconstructed Higgs boson (reconstructed Y) candidate is calculated.  The candidates are considered correctly paired if $\Delta R< 0.25$.
The efficiency ranges from 77.9 to 99.5\%, depending on the signal mass hypothesis, where the efficiency increases as the difference $\mXreco-\mYreco$ becomes larger.

For both the 3\PQb and 4\PQb samples, events are subdivided into analysis regions that depend on the reconstructed mass of the Higgs boson candidate (\mHreco). The signal region (SR) is defined by  $|\mHreco - 125| < 20\GeV$, the validation region (VR) by $20 < |\mHreco - 125| < 30\GeV$, and control region (CR) by $30 < \abs{\mHreco - 125} < 60\GeV$.
The search for signal is performed in the SR of the 4\PQb sample, SR(4\PQb).

After the analysis regions are defined, a kinematic constraint is imposed on the \PQb~jets forming the \PH candidate.  The $\eta$, $\phi$, and \pt of the \PQb~jets are adjusted within their uncertainties to enforce the mass of the \PH candidate to be 125\GeV. The \PX candidate is then reconstructed with the updated \PH candidate. This technique is described in Ref.~\cite{DHondt:2006iej}.

\section{Background modeling}
\label{sec:bkgModeling}

The search for a signal in SR(4\PQb) requires an estimation of the expected background in that region. The background composition is dominated by QCD multijet ($\sim$90\%) and \ttbar ($\sim$10\%) events, with the relative contributions estimated using the simulated samples discussed in Section~\ref{sec:sims}. The modeling of the background is done using control samples in data, based on the two-dimensional distribution of 3\PQb events in the (\mXreco,~\mYreco) space. This approach inclusively accounts for all processes that contribute to the background. The background modeling is performed separately for all three data-taking years to account for the different event selection criteria and trigger algorithms.

A BDT-based reweighting method, described in Ref.~\cite{Rogozhnikov:2016bdp}, is used to weight events in the background-dominated 3\PQb sample to effectively model the differences between the 3\PQb and 4\PQb samples as a function of multiple observables.

The BDT is trained with collision data in the CR(3\PQb) and CR(4\PQb).
The obtained weights take into account both residual differences in the event yield normalization and the differences in the shapes of the kinematic distributions between the 4\PQb and 3\PQb samples.
The ratio of the event yields of the 3\PQb to 4\PQb samples ranges between 2.19 to 2.55, depending on data-taking year. The differences in the yields for each year depend on the trigger algorithms and the \PQb~tagging efficiencies.

Variables used as input to the BDT are: \mXreco, \mYreco, the \pt of all four selected \PQb~jet candidates, the \pt and $\eta$
of both the reconstructed H (\Hreco) and reconstructed Y (\Yreco) candidates,
the angular separation $\Delta R$ between the two jets that form \Hreco,
and $\Delta R$ between the two jets that form \Yreco.
Additionally, the orthogonal distance of the event in the (\mXreco,~\mYreco) plane from the kinematic boundary defined by  $\mXreco - \mYreco=125\GeV$ is included as an input variable. This variable is used to provide correlated \mXreco and \mYreco information to the BDT during training.

To optimize the training parameters of the BDT model, data in the CR(4\PQb) are compared with the data in CR(3\PQb) after the reweighting is applied. This step makes use of the Kolmogorov--Smirnov~(KS) distance test~\cite{kstest}, as well as a dedicated BDT discriminator trained to separate the two samples. After optimization, both techniques indicate that the reweighting model performs well. The KS metrics for each reweighted variable decrease, approaching zero, and the areas under the receiver operating characteristic curves of the discriminator are equal to 0.5, indicating the reweighted sample can no longer be distinguished from the target sample.

As a consequence of the \PQb jet pairing strategy and the definition of the analysis regions, the CR does not contain data in the range $95<\mYreco<155\GeV$. However, the BDT reweighting method is designed to account for variations in the event yields between the 3\PQb and 4\PQb samples across the analysis regions. To verify this, a statistically independent sample is defined. Data outside the SR were re-analysed with the same analysis procedures described in this document, but assuming an \mH of 205\GeV. In this dataset, the signal region, indicated as SR$_{205}$, is not blinded and the background performance is validated. With this sample, good agreement between the expected and observed background distributions is observed with bin-by-bin disagreements up to around 10\%.

\begin{figure*}[htb!]
\centering
{\includegraphics[width=0.47\textwidth] {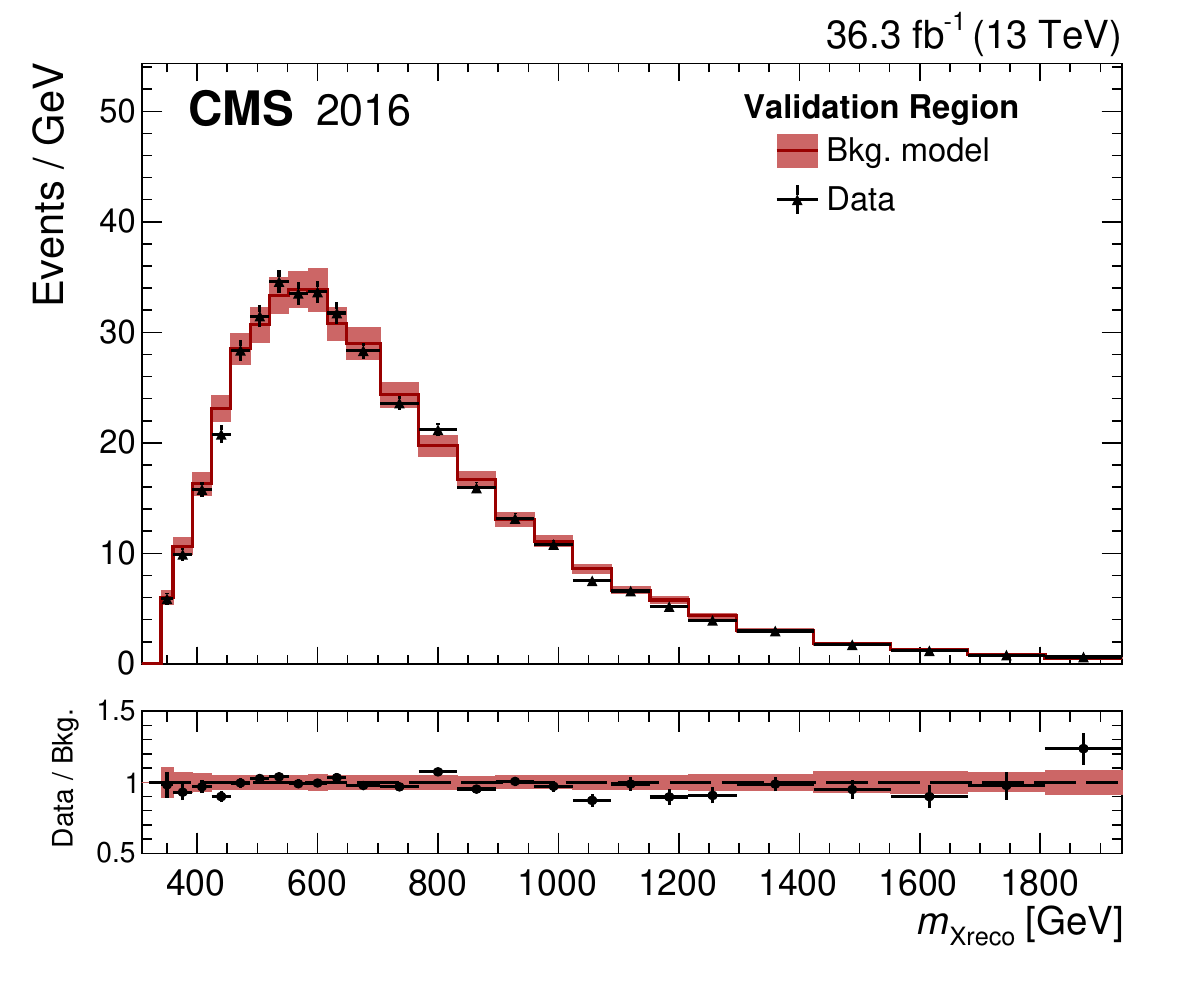}}
{\includegraphics[width=0.47\textwidth] {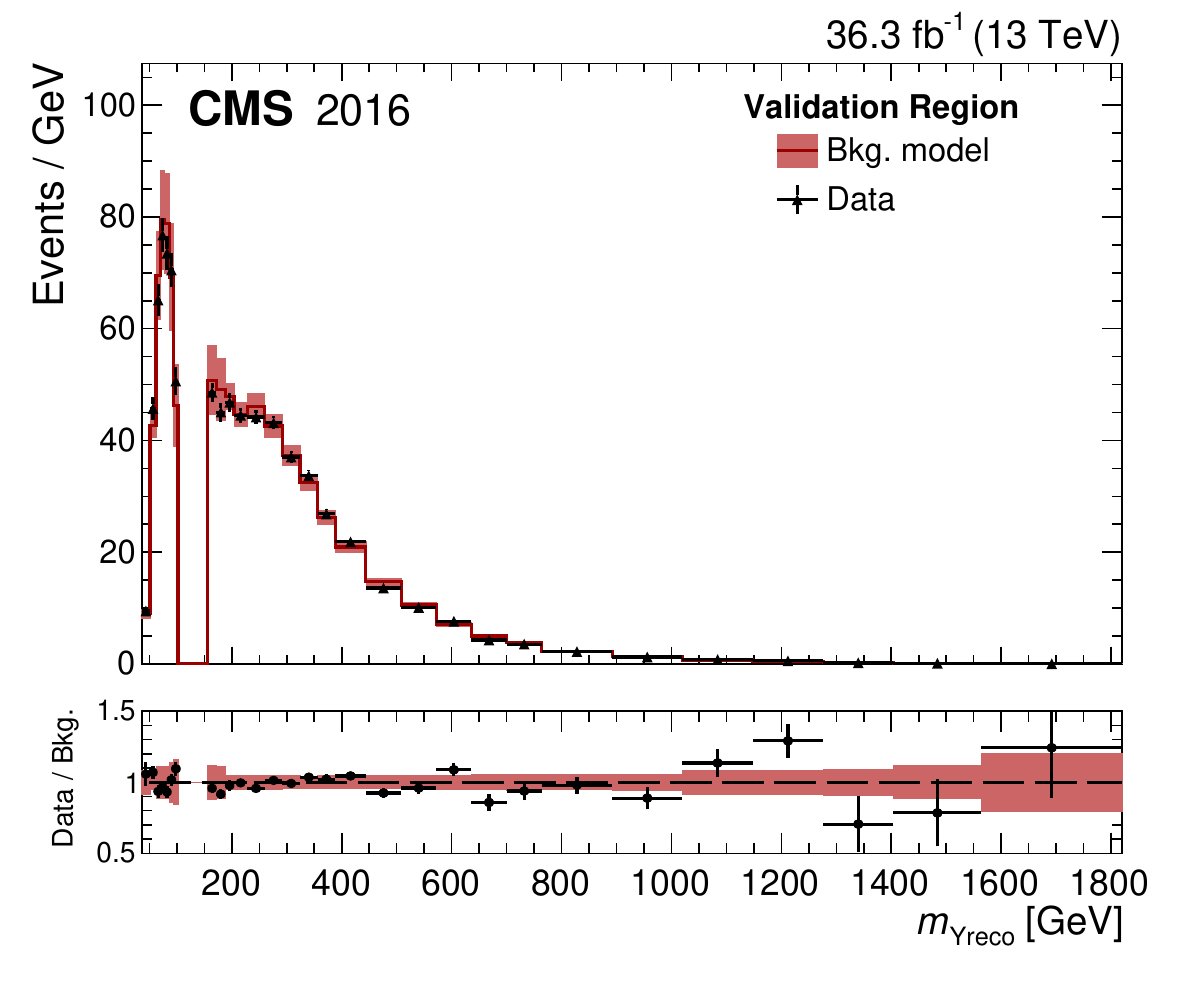}}\\
{\includegraphics[width=0.47\textwidth] {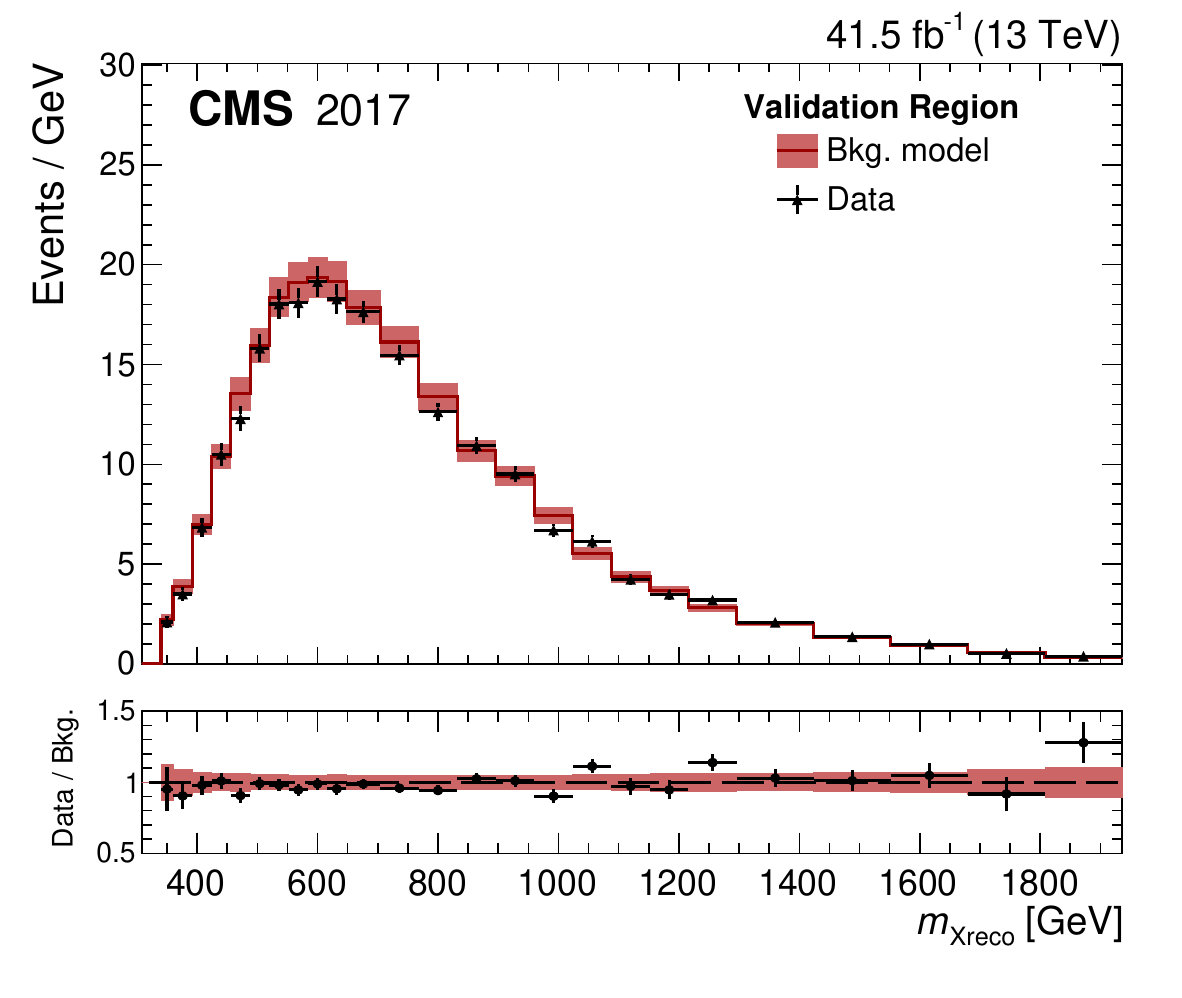}}
{\includegraphics[width=0.47\textwidth] {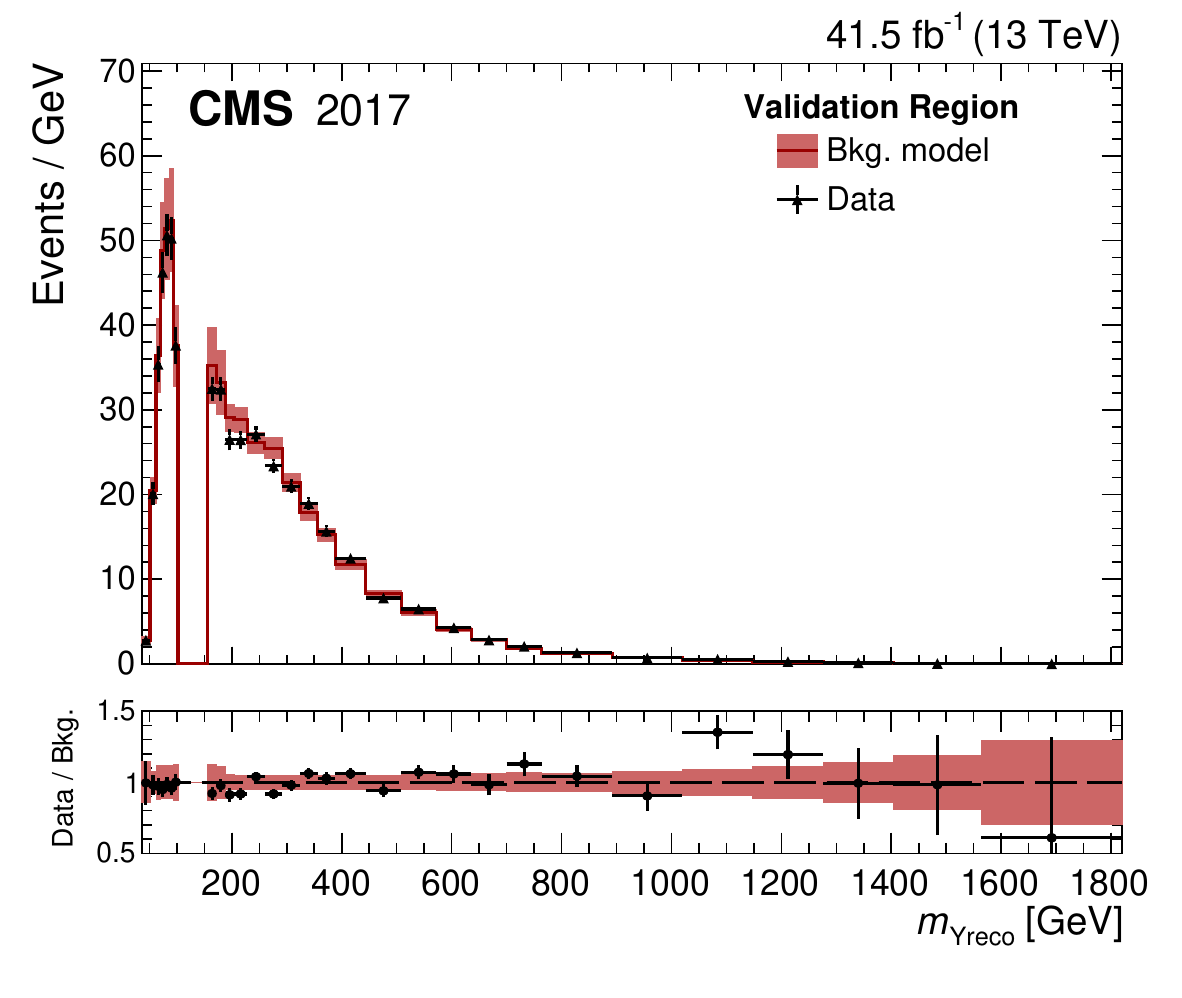}}\\
{\includegraphics[width=0.47\textwidth] {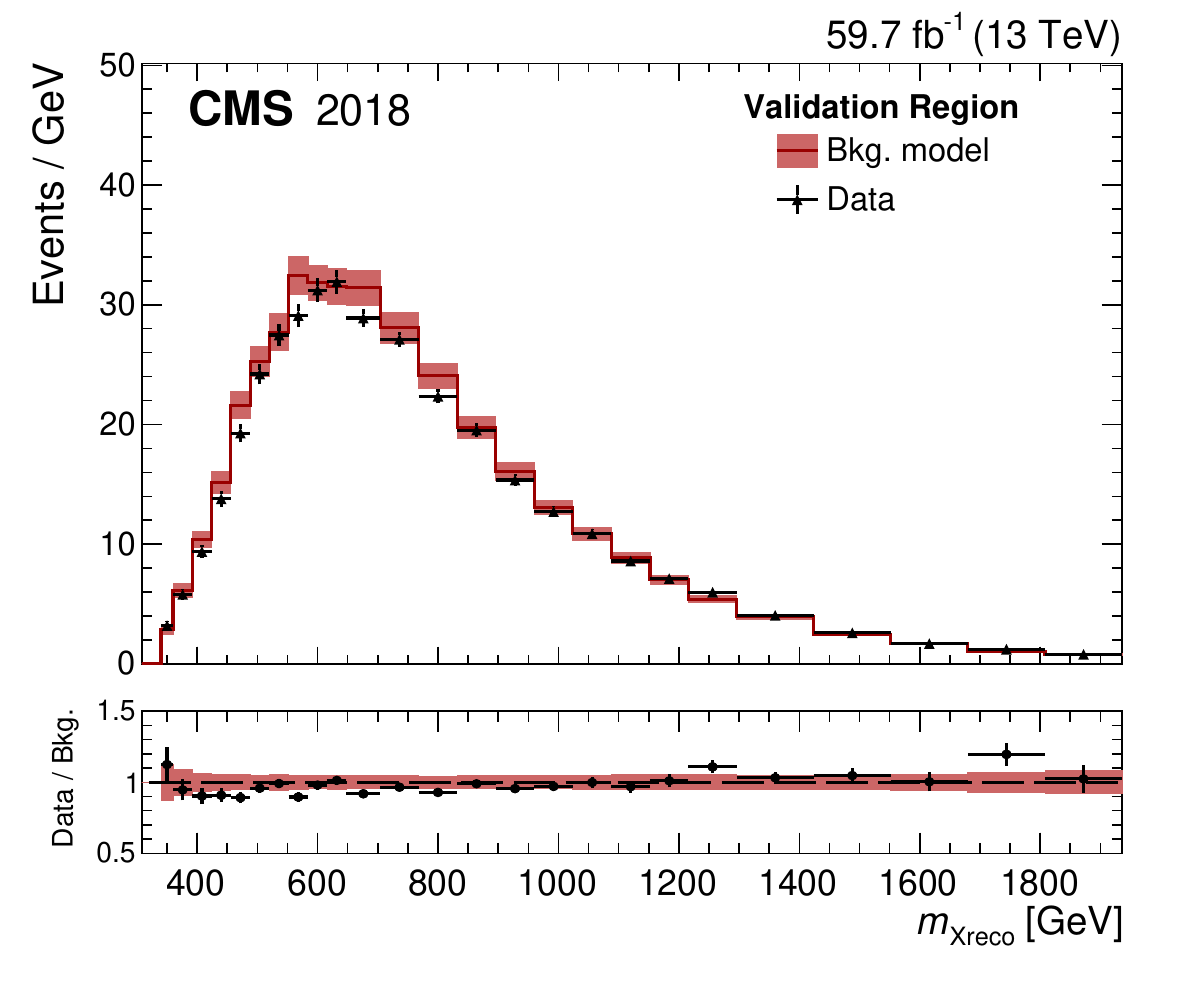}}
{\includegraphics[width=0.47\textwidth] {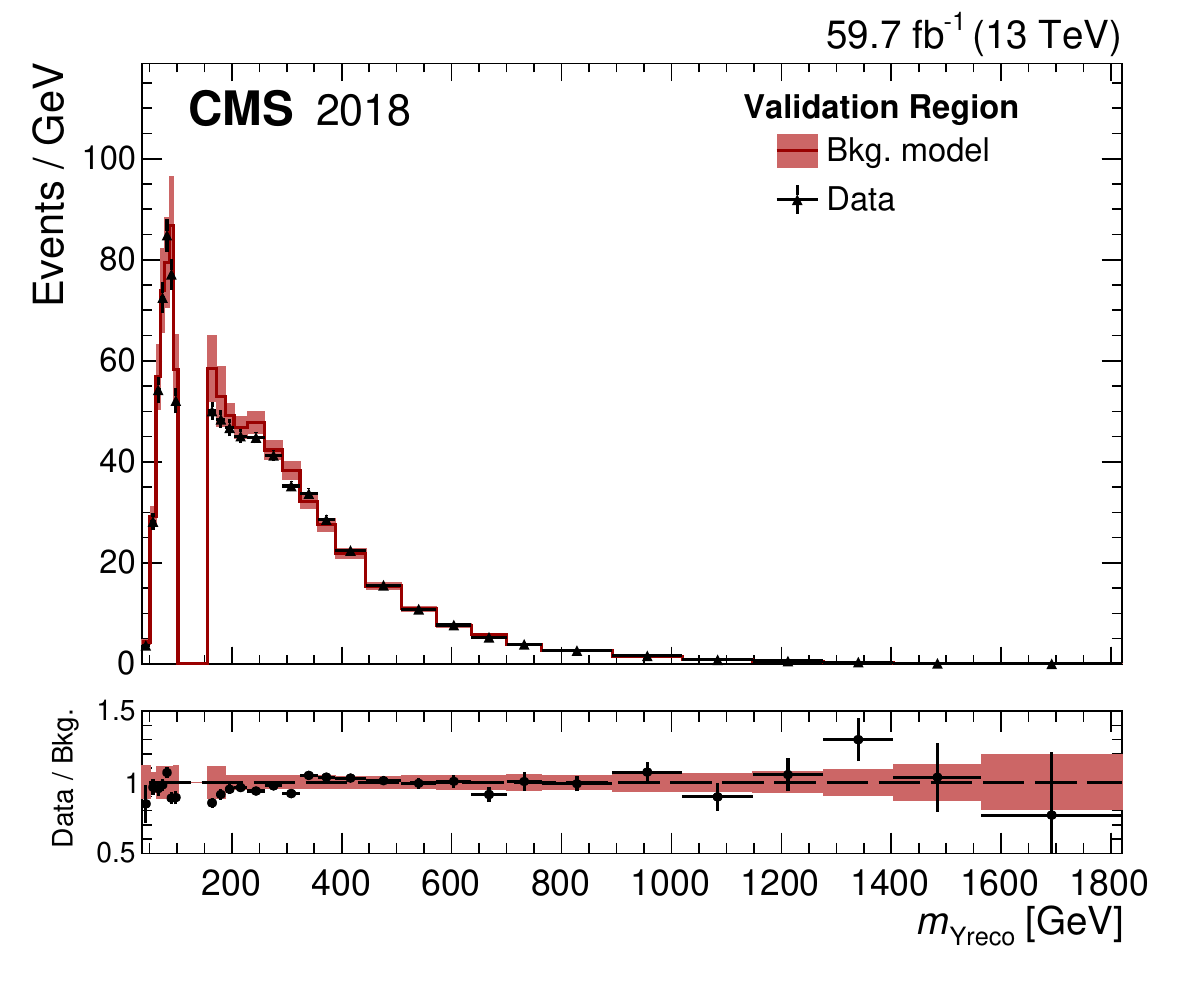}}
\caption{\label{fig:mXmYvarsVR}
    Event distributions in the validation regions for \mXreco (left column) and \mYreco (right column), shown separately for the three data-taking years: 2016, 2017, and 2018 (upper, middle, and lower rows, respectively).
    The VR(4\PQb) data are shown in black and BDT reweighted VR(3\PQb) model in red. The uncertainties include the statistical component added in quadrature with the uncertainties described in Section~\ref{sec:syst}. The ratios of VR(4\PQb) to VR(3\PQb) (target over model) are in the lower panels. The ratio of the VR(3\PQb) model uncertainty to the central value of the model is shown with the red band in the lower panel.
}
\end{figure*}

Validation of the background model is performed by applying the derived weights to the VR(3\PQb) and comparing the kinematic distributions and yields between the reweighted VR(3\PQb) and the VR(4\PQb). It is observed that the kinematic distributions and yields are the same within the derived uncertainties. Table~\ref{tab:yields} shows the event yield per analysis region, including the estimated event yields extracted from the background modeling. Comparisons of the \mXreco and \mYreco variables in the validation regions are shown in Fig.~\ref{fig:mXmYvarsVR}, for each data-taking year.

\begin{table}[h!]
\topcaption{\label{tab:yields} Summary of the number of events in the samples and analysis regions for each data-taking year.
The last three rows provide the estimated event yields derived from the background model (EST) and can be compared to those in the 4\PQb dataset. Systematics uncertainties associated with these estimates are discussed in Section~\ref{sec:syst}.
}
\centering
\begin{scotch}{l  r  r  r }
Dataset                      & 2016         & 2017          & 2018          \\
\hline
$\mathrm{CR(3b)}$       &	73408       &	39997       &	80449       \\
$\mathrm{VR(3b)}$       &	44048       &	22073       &	44863       \\
$\mathrm{SR(3b)}$       & 129763       &	59957       &  122054       \\[\cmsTabSkip]
$\mathrm{CR(4b)}$       &	28838       &	18295       &	31520       \\
$\mathrm{VR(4b)}$       &	17283       &  	10313       &	17551       \\
$\mathrm{SR(4b)}$       &  50301       &   27042       &   47236       \\[\cmsTabSkip]
$\mathrm{CR(4b)^{EST}}$   &  28236    &	18213    &	31322    \\
$\mathrm{VR(4b)^{EST}}$   &  17551    &	10508    &	18164    \\
$\mathrm{SR(4b)^{EST}}$   &  52179    &	28864    &  49954    \\
\end{scotch}
\end{table}

An additional test is performed to verify whether a potential signal contamination of the 3\PQb sample induces a bias in the background determination.
The test uses the background model in the SR (derived by reweighting SR(3\PQb)) data, and simulated signal.
Best fit signal strengths are extracted using the simulated signal and either the background model or a pseudo-data sample constructed using the background model and injected signal contamination.
This study is repeated for a subset of the mass hypotheses considered for the search and for injected signal strengths up to five times the expected sensitivity of this analysis.
In all cases, the results obtained with and without the injected signal contamination are found to be compatible within one standard deviation. These results indicate that a potential self-bias effect of the background model does not impact the final result.

\begin{figure*}[htb!]
\centering
{\includegraphics[width=0.47\textwidth] {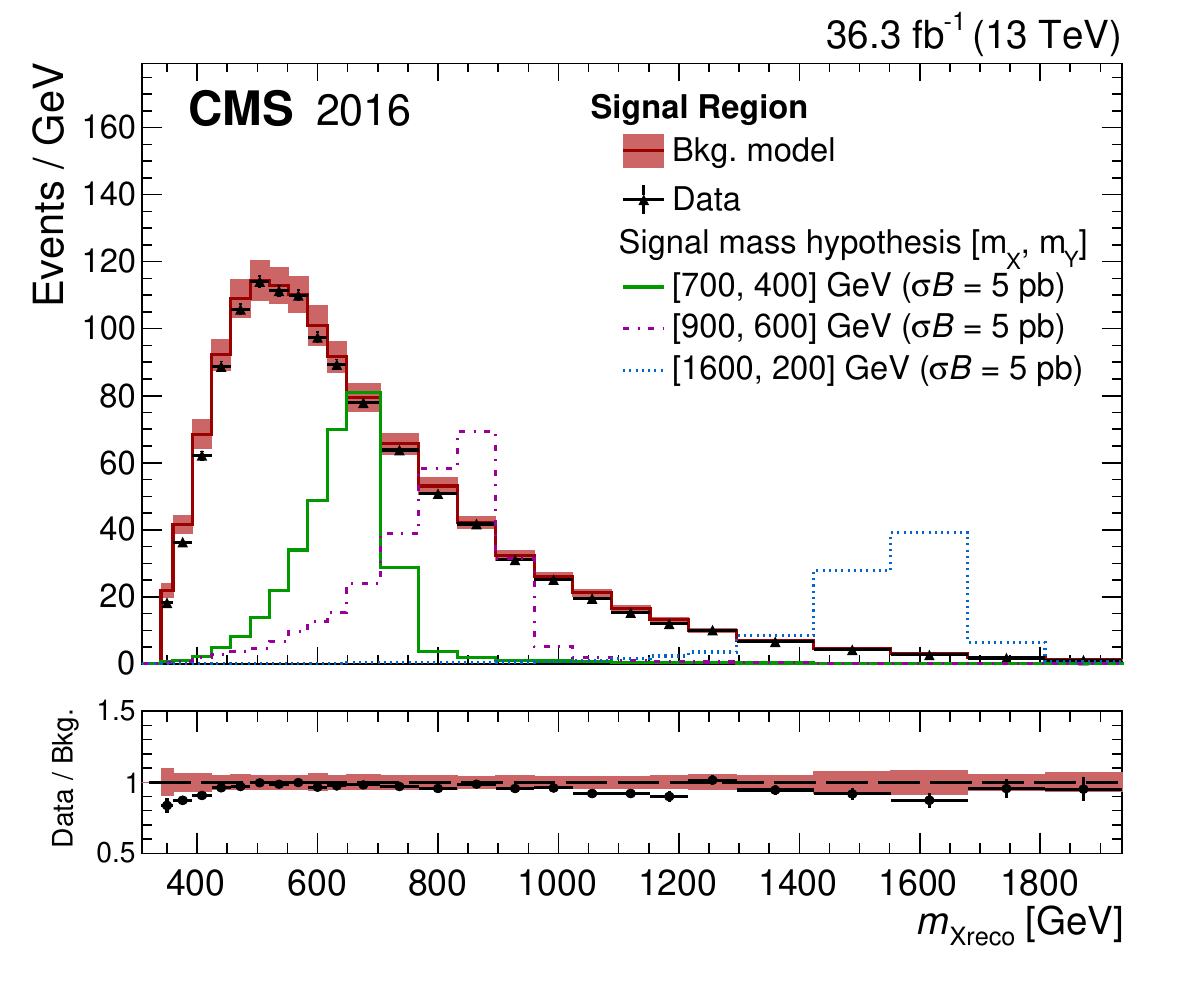}}
{\includegraphics[width=0.47\textwidth] {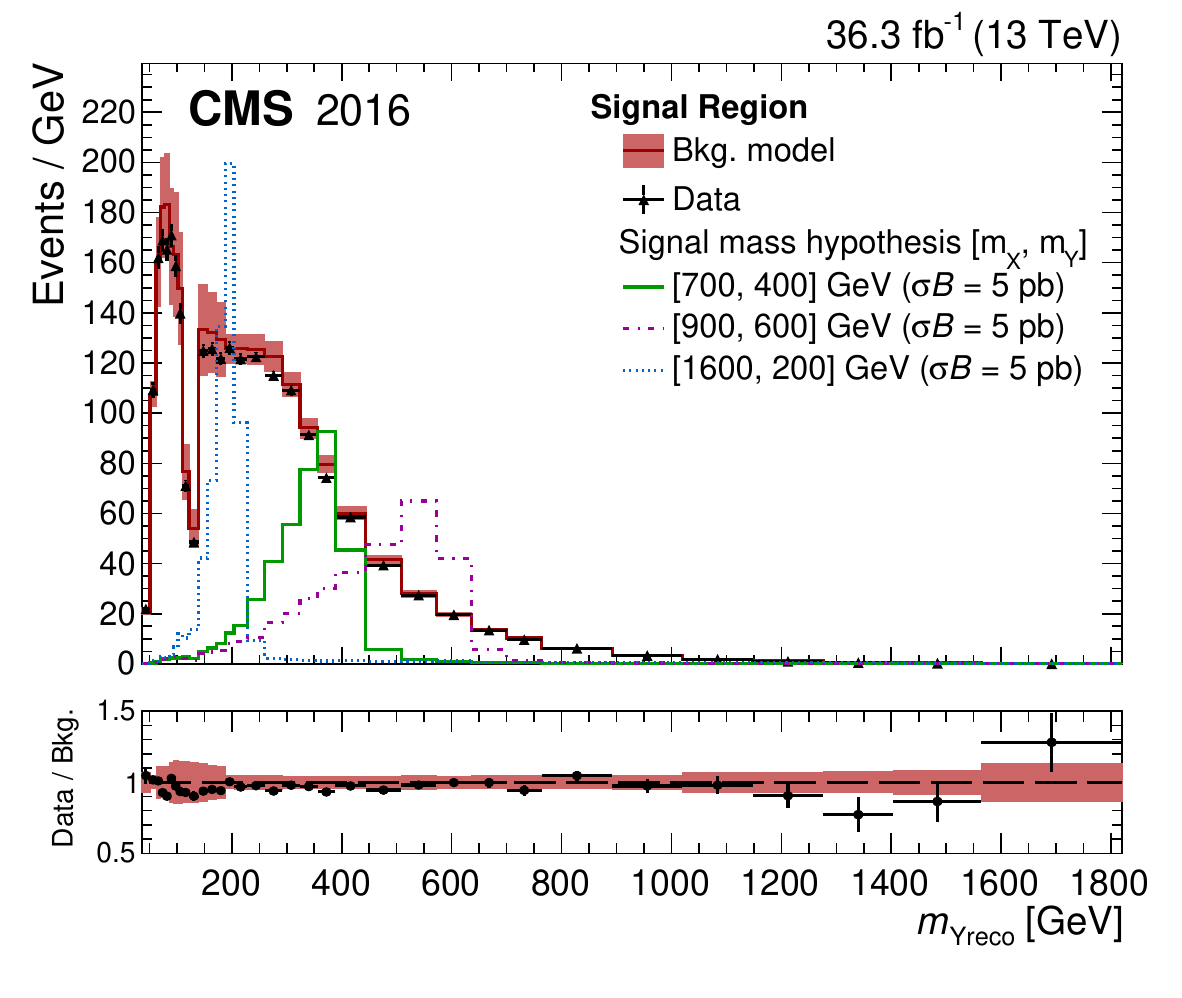}}\\
{\includegraphics[width=0.47\textwidth] {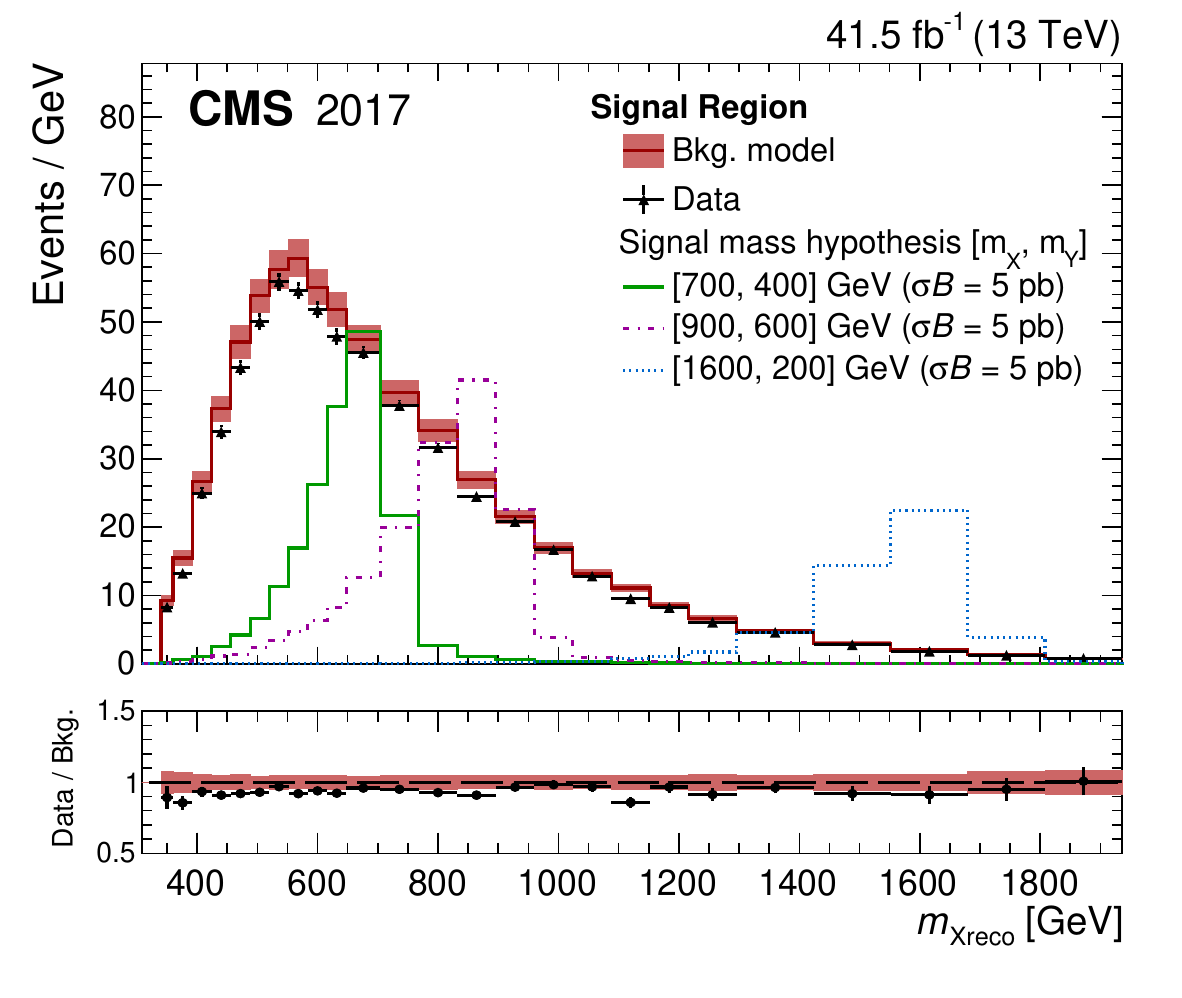}}
{\includegraphics[width=0.47\textwidth] {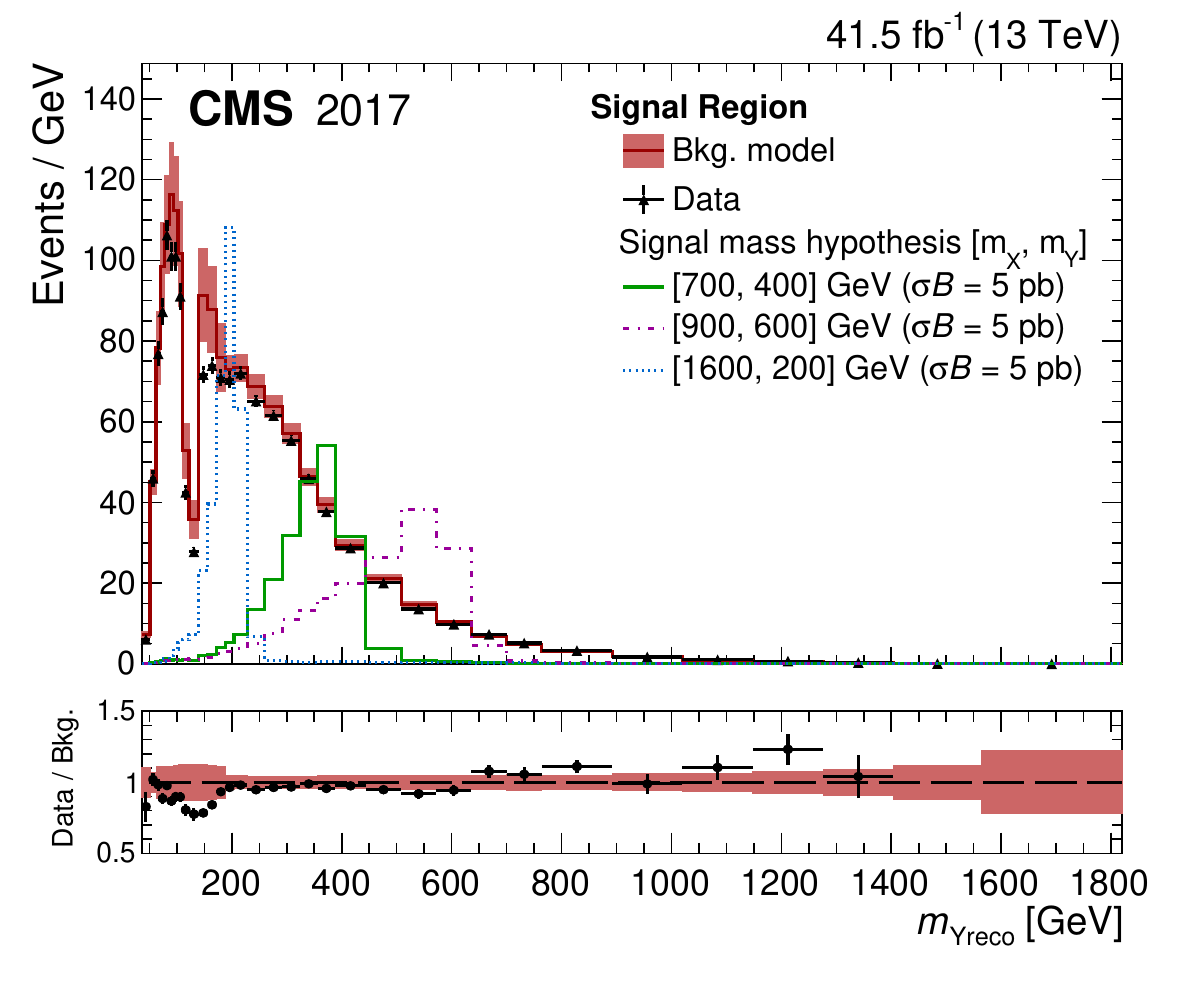}}\\
{\includegraphics[width=0.47\textwidth] {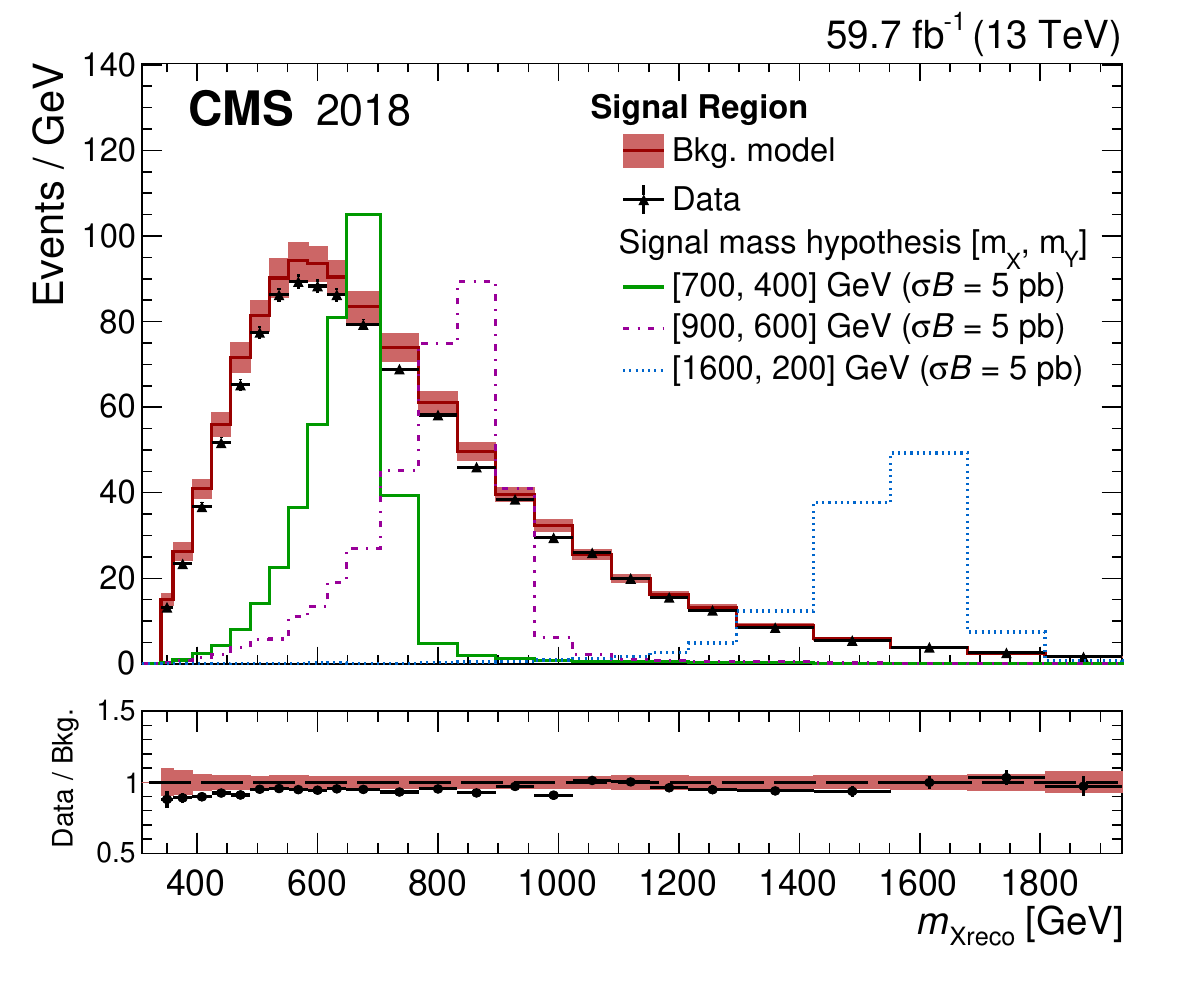}}
{\includegraphics[width=0.47\textwidth] {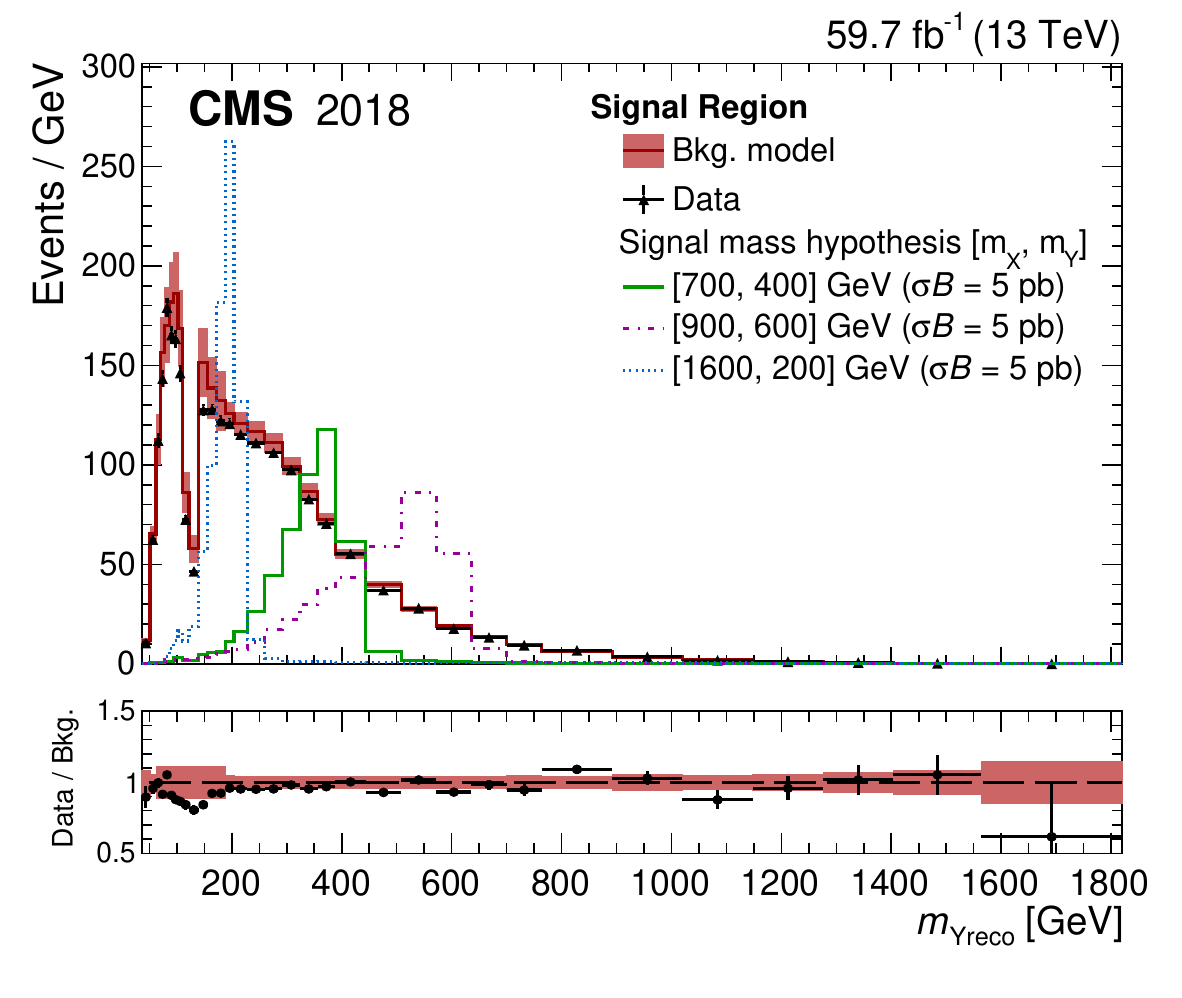}}
\caption{\label{fig:mXmYvarsSR}
    Event distributions in the signal region for \mXreco (left column) and \mYreco (right column), shown separately for the three data-taking years: 2016, 2017, and 2018 (upper, middle, and lower rows, respectively).
    The SR(4\PQb) data are shown in black and BDT reweighted SR(3\PQb) model in red. The uncertainties include the statistical component added in quadrature with the uncertainties described in Section~\ref{sec:syst}. Three selected signal mass hypotheses are overlaid. The signal histograms are scaled to have $\sigma \mathcal{B}$ values of 5\unit{pb}. The ratios of SR(4\PQb) to SR(3\PQb) (target over model) are in the lower panels. The ratio of the SR(3\PQb) model uncertainty to the central value of the model is shown with the red band in the lower panel.
}
\end{figure*}

Distributions of the most important analysis observables, \mXreco and \mYreco, for the SR(4\PQb) and reweighted SR(3\PQb) are shown in Fig.~\ref{fig:mXmYvarsSR},
including the systematic uncertainties described in Section~\ref{sec:syst}.
The data are found to be in good agreement with the background model, and no localized excesses or shape deviations are observed.
The lower number of events in the \mYreco distribution near the Higgs boson mass is due to the candidate reconstruction method, which identifies the \PQb~jet pair with an invariant mass closest to 125\GeV as the H candidate.

\section{Systematic uncertainties}
\label{sec:syst}
Systematic uncertainties are introduced by two components of the analysis: the simulated signal and the background estimate, which relies on control samples in data. Systematic uncertainties in the signal expectations arise from discrepancies between simulation and data, due in part to imperfect modeling of the detector response and the limited number of events in the simulated samples.
These correspond to uncertainties that arise from the trigger efficiency, \PQb~tagging efficiency, \PQb~tagging energy regression smearing, jet energy scale, jet energy resolution, pileup reweighting, PDF, parton shower, and integrated luminosity.
The dominant sources of uncertainty for the signal hypotheses arise from the trigger and \PQb~tagging efficiencies.
The trigger efficiency is the largest contributor to the total uncertainty for signals close to the kinematic boundary in the (\mXreco,~\mYreco) plane, defined by  $\mXreco - \mYreco=125\GeV$.

The difference between the trigger efficiency in data and simulated events, referred to as a scale factor, is determined by measuring the efficiency in a \ttbar-dominated data sample and the corresponding simulated sample.
The uncertainties in the trigger scale factor are implemented as the trigger efficiency systematic uncertainty for the signal events.
This component of the uncertainty is higher in the region shortly before the efficiency plateaus to its maximum, leading to the larger impact along the $\mXreco - \mYreco=125\GeV$ kinematic boundary in the (\mXreco,~\mYreco) plane.
Inaccuracy in the trigger efficiency measurements contributes up to 30\% of the total uncertainty, depending on the signal and data-taking year.
The uncertainties in the \PQb~tagging \textsc{DeepJet} discriminants are accounted for following Ref.~\cite{Bols:2020bkb}. This uncertainty contributes up to 30\% of the total uncertainty, with variations based on the signal and the corresponding data-taking year.
Jet energy scale and resolution uncertainties are taken into account by varying the jet properties within their uncertainties.
A shape uncertainty associated with the \PQb~jet energy regression smearing is determined by shifting the jet resolution up and down by 10\% and rerunning the analysis procedures.
The uncertainty in the pileup model is determined by varying the inelastic pp cross section by 4.6\%~\cite{CMS:2018mlc}.
The PDF uncertainties are based on the estimation performed in Ref.~\cite{hh4b} and range from 1 to~3\%, increasing linearly as a function of the  \mXreco value of the signal. The parton shower uncertainty of about 7\% is derived as in Ref.~\cite{CMS:2022cpr}.
The integrated luminosities for the 2016, 2017, and 2018 data-taking years have 1.2--2.5\% individual uncertainties~\cite{CMS-LUM-17-003,CMS-PAS-LUM-17-004,CMS-PAS-LUM-18-002}, with an overall uncertainty of 1.6\%. The systematic uncertainties are treated as uncorrelated across the data-taking years, except for the PDF and parton shower uncertainties.

Given that the search covers a large range of masses, the uncertainty on the background estimation depends on the mass hypothesis being considered.
The dominant source of background systematic uncertainty arises from the limited number of data events in the SR(3\PQb) region. These uncertainties are uncorrelated across the individual bins of the background model, and propagated as bin-by-bin Poisson uncertainties from the SR(3\PQb) to the SR(4\PQb). This source of uncertainty contributes up to 95\% of the total background uncertainty, depending on the signal and data-taking year. The uncertainty on the 4\PQb-to-3\PQb normalization factor is determined by comparing the yields in the 4\PQb-to-3b VRs and contributes at most 25\% of the total uncertainty for any given signal. The potential dependence of the background estimate on the selected boundaries of the CRs is determined by measuring the background in alternative CRs ($35 < \abs{\mHreco - 125} < 55\GeV$  and $25 < \abs{\mHreco - 125} < 65\GeV$). The derived systematic uncertainty contributes less than 15\% of the total uncertainty for all signal hypotheses.
Finally, a 10\% yield discrepancy for the bins in the range $62<\mYreco<105\GeV$ and $145<\mYreco<188\GeV$ was observed between VR(4\PQb) and VR(3\PQb).
This is implemented as a systematic uncertainty for the bins in the range $62<\mYreco<188\GeV$ in the SR.
The value of this systematic is consistent with the disagreements observed in SR$_{205}$.
This uncertainty is included in the VR plots of Fig.~\ref{fig:mXmYvarsVR}.

\section{Results}
\label{sec:results}
\begin{figure*}[htbp!]
\centering
{\includegraphics[width=0.49\textwidth] {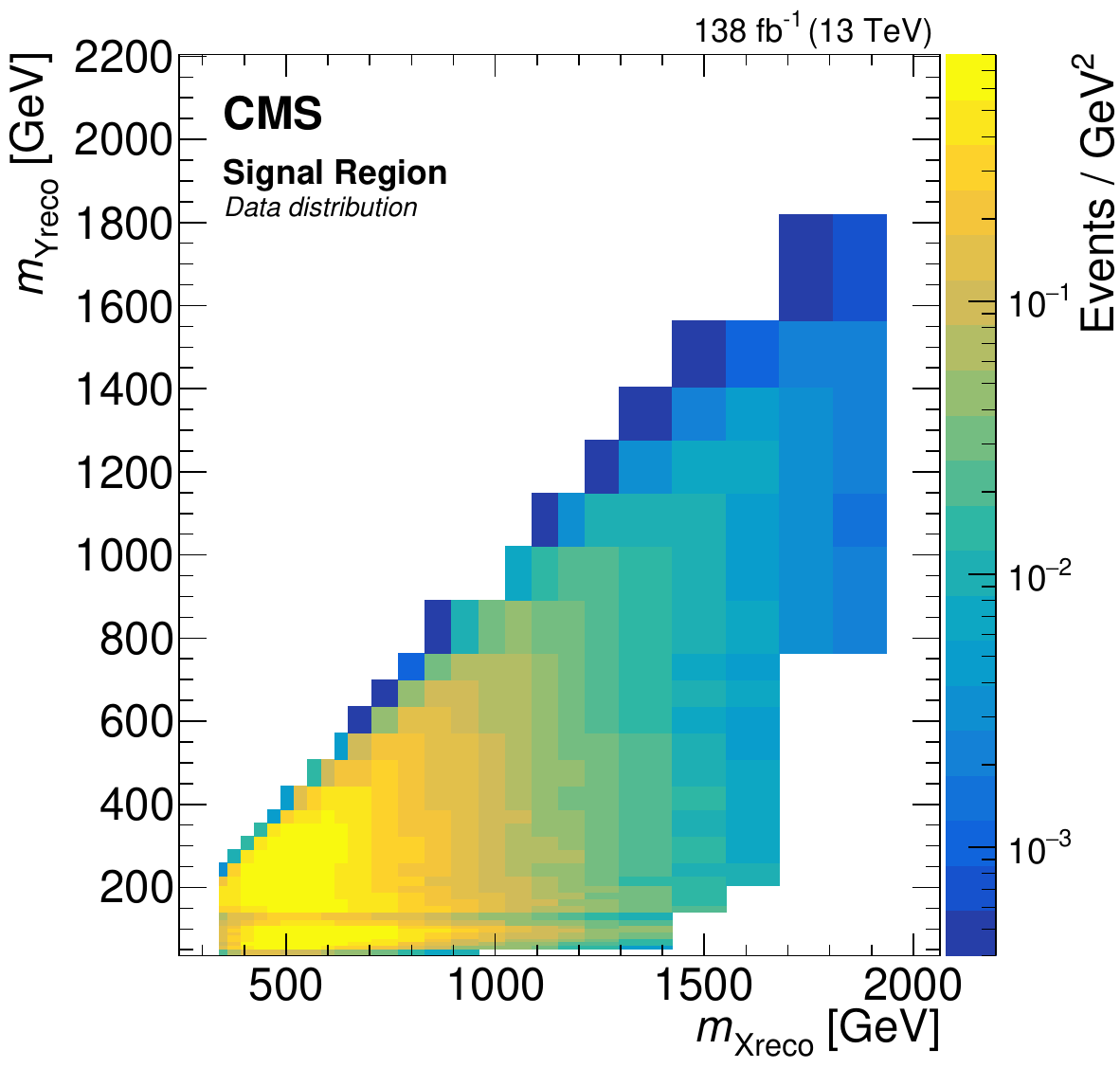}}\hspace{\fill}
{\includegraphics[width=0.49\textwidth] {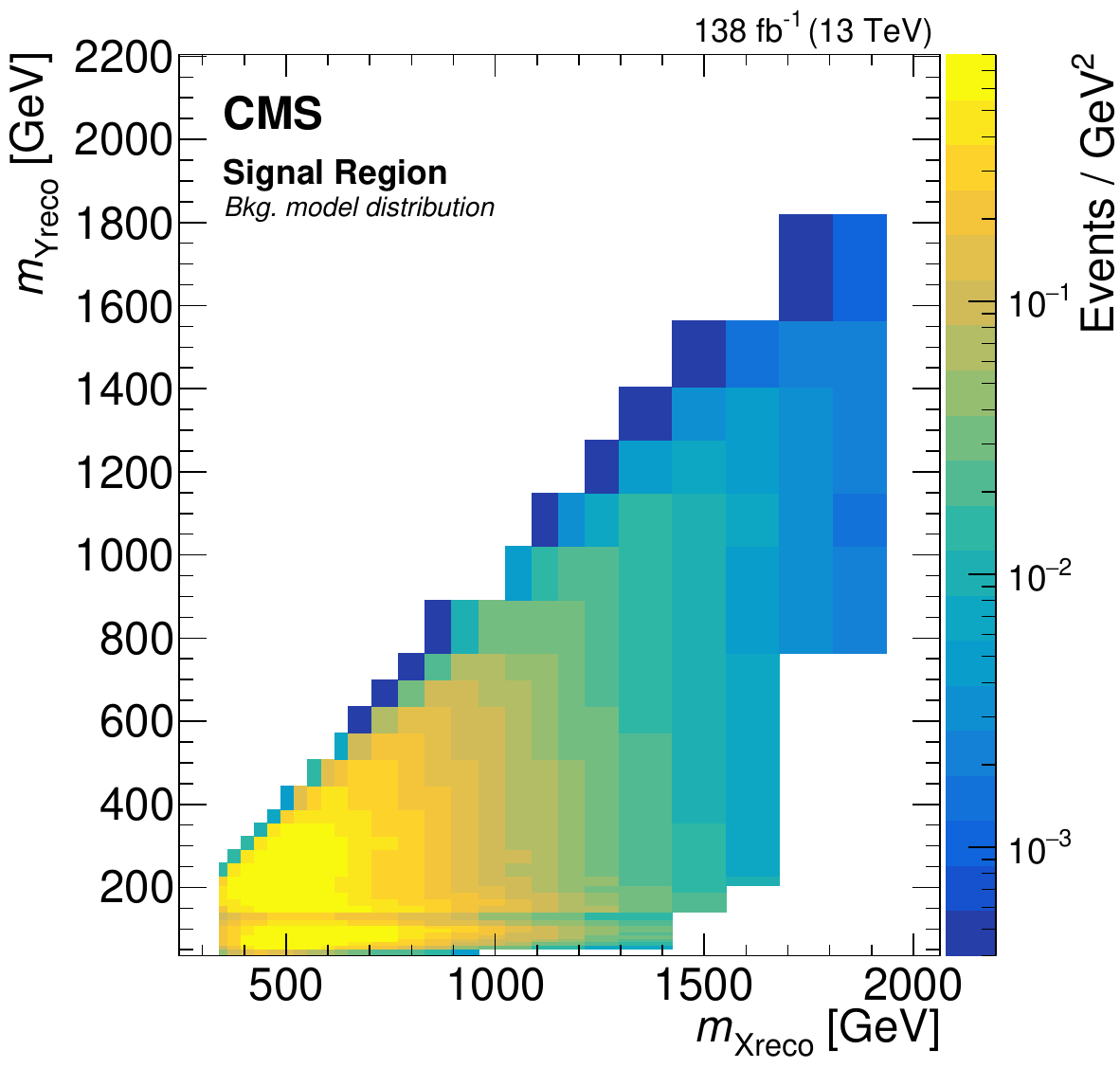}}\\
{\includegraphics[width=0.49\textwidth] {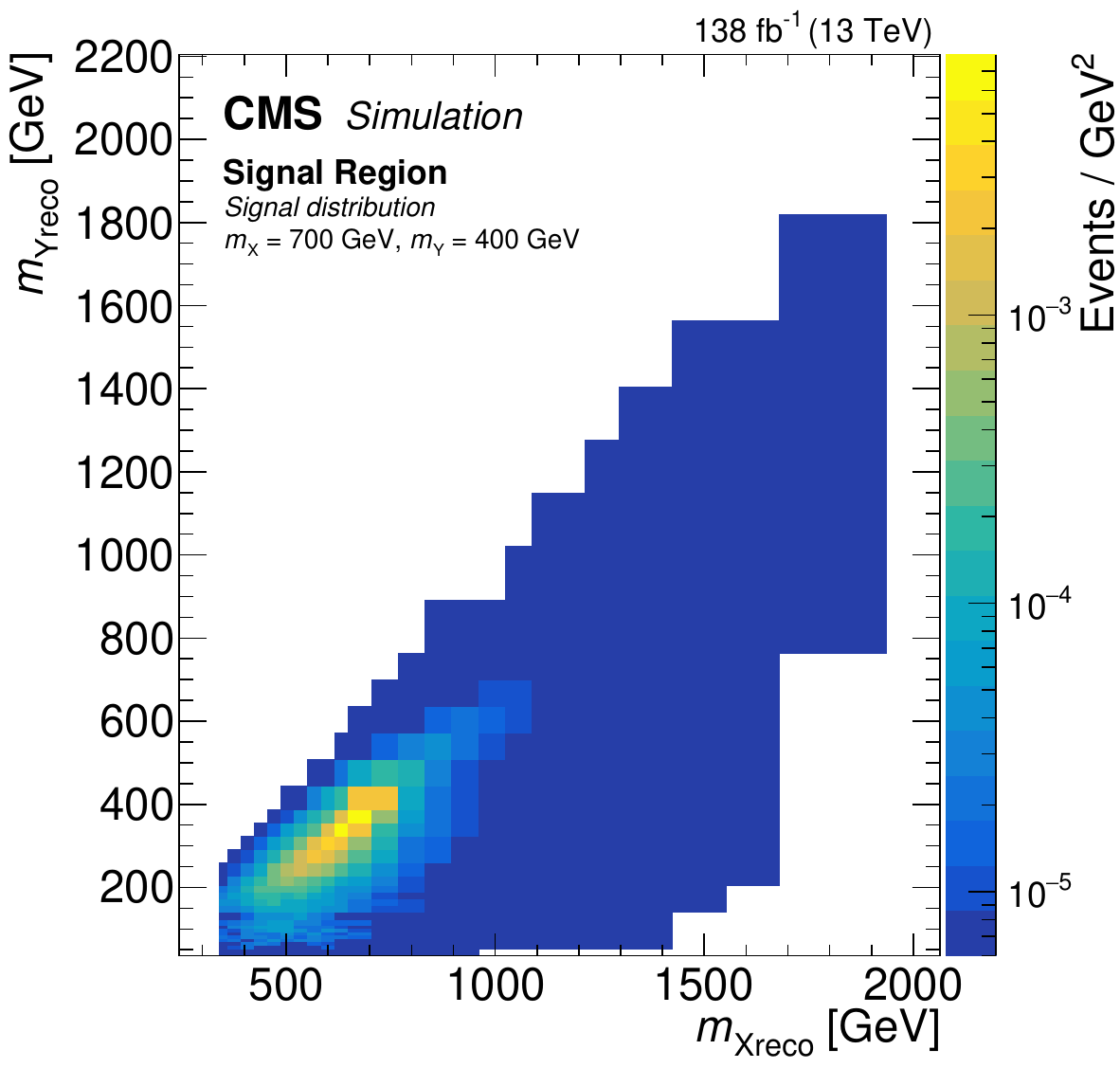}}
\caption{\label{fig:dataMCplots}
    Distributions of the events in the (\mXreco,~\mYreco) plane observed in the SR(4\PQb).
The upper plots show events in data (left) and the background model (right). The lower plot shows the distribution of events for the signal hypothesis corresponding to \mX~=~700\GeV and \mY~=~400\GeV. In each plot, there are empty bins in the high-\mXreco and low-\mYreco region. These areas have been excluded because the events have highly boosted topology.
}
\end{figure*}

The signal is extracted through a two-dimensional fit in the (\mXreco,~\mYreco) plane. Distributions of events in the SR(4\PQb) for data, background, and one of the signal mass hypotheses are shown for the sum of all data-taking years in Fig.~\ref{fig:dataMCplots}.
The boundaries of the bins in Fig.~\ref{fig:dataMCplots} are chosen taking into account the \mXreco and \mYreco detector resolutions. In the high mass regions, bins are made to have larger areas to reduce the systematic impact due to the low statistics in the 3\PQb dataset.
To minimize the impact of potential mismodeling of the trigger efficiency and jet reconstruction, events with $\mXreco - \mYreco < 120\GeV$ are excluded from the statistical interpretation of results. Regions with high \mXreco and low \mYreco are not considered, since the analysis is not optimized for the decay of  highly Lorentz-boosted Y resonances.
The boundaries in the (\mXreco,~\mYreco) plane that are used for the search range from 340~to~1936\GeV in \mXreco.
The minimal value for \mYreco is set at 36\GeV.
For the larger \mXreco values, the \mYreco minimum boundary is increased as: [$\mXreco > 959$, $\mYreco > 52\GeV$],
[$\mXreco > 1200$, $\mYreco > 80\GeV$],
[$\mXreco > 1423$, $\mYreco > 141\GeV$],
[$\mXreco > 1550$, $\mYreco > 205\GeV$],
and [$\mXreco > 1678$, $\mYreco > 770\GeV$].
Similar to the discussion regarding Fig.~\ref{fig:mXmYvarsSR}, a lower number of events is observed in the \mYreco region around the Higgs boson mass.

\begin{figure*}[htbp!]
\centering
{\includegraphics[width=0.95\textwidth] {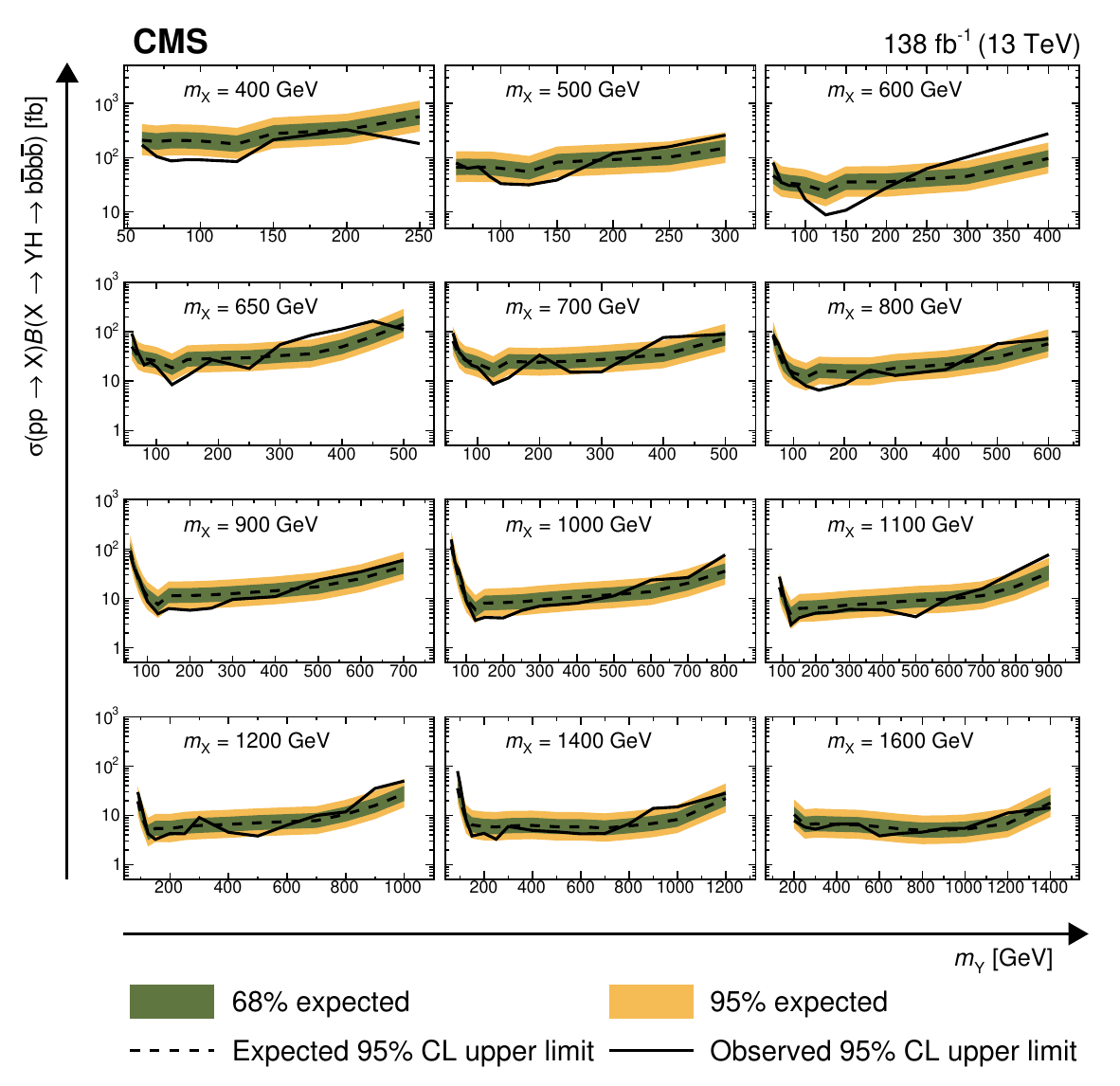}}
\caption{\label{fig:limits}
    Expected and observed 95\% CL upper limits on $\sigma(\Pp\Pp \to \PX) \mathcal{B} (\XHY)$. The limits are shown as functions of \mYreco for selected values of \mXreco. The black dashed and solid lines represent expected and observed limits, respectively. The dark green and light yellow bands represent the $\pm 1$ and $\pm 2$ standard deviations for the expected limit, respectively. The largest excess (deficit) of the observed limit compared with the expected limit is for $\mXreco = 600\GeV$ and $\mYreco = 400\GeV$ ($\mXreco = 600\GeV$ and $\mYreco = 150\GeV$).
}
\end{figure*}

\begin{figure}[htb!]
\centering
{\includegraphics[width=\cmsFigWidth] {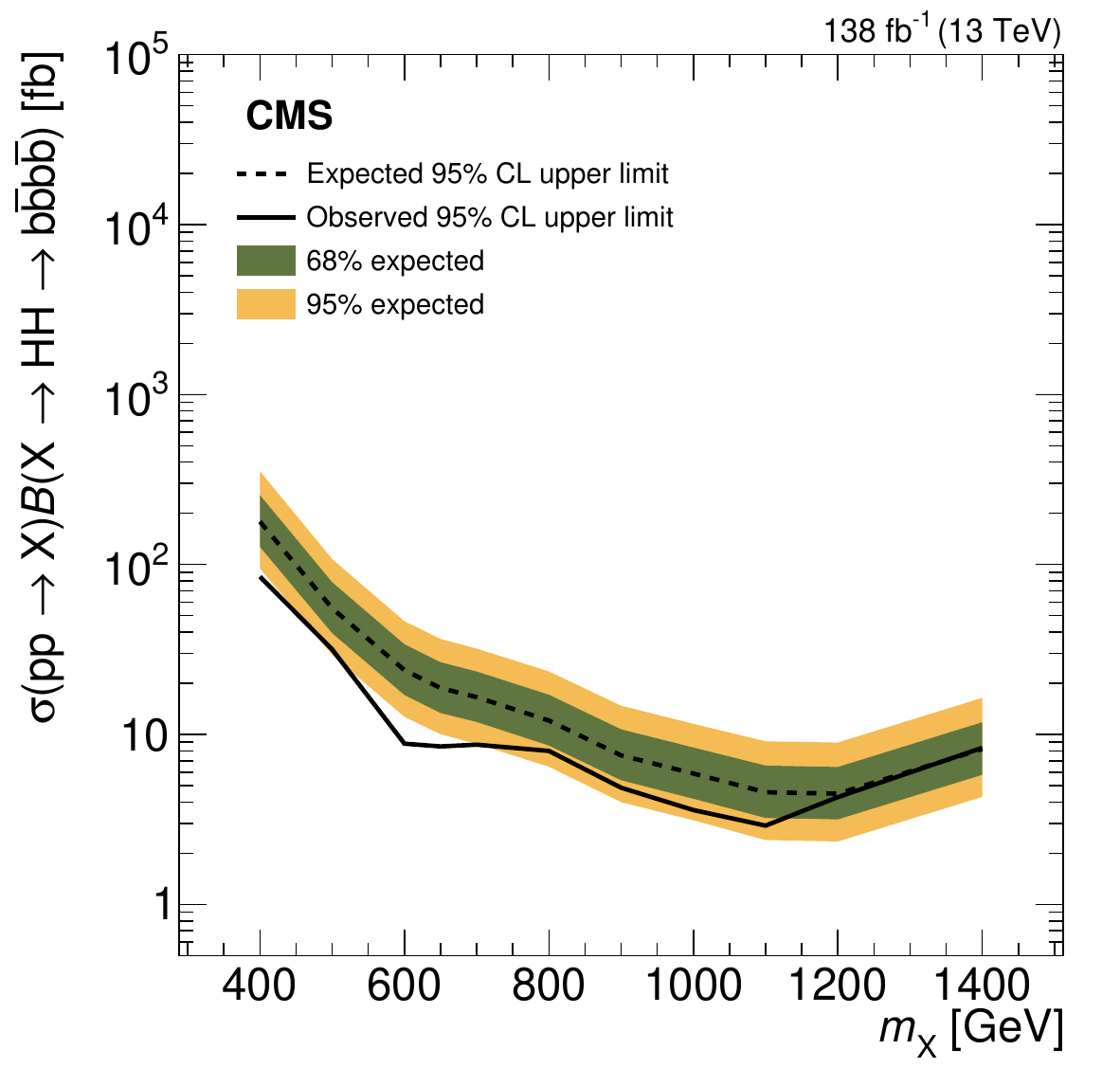}}
\caption{\label{fig:HHlimits}
    Expected and observed 95\% CL upper limits on $\sigma(\textrm{pp} \to \PX) \mathcal{B} (\XHH)$. The black dashed and solid lines represent expected and observed limits, respectively. The dark green and light yellow bands represent the $\pm$1 and $\pm$2 standard deviations for the expected limit, respectively.
}
\end{figure}

A binned maximum likelihood fit is performed, using the previously discussed systematic uncertainties as nuisance parameters. The prior distribution used for each nuisance parameter is either Gaussian (for shape uncertainties) or log-normal (for normalization uncertainties).
The following results have been determined using the statistical analysis tool \textsc{Combine}~\cite{CMS:2024onh}, which is based on the \textsc{RooFit}~\cite{Verkerke:2003ir} and \textsc{RooStats}~\cite{Moneta:2010pm} frameworks.
The observed data and the background are found to be in agreement with each other.
The observed data are used to set 95\% confidence level (CL) upper limits on the cross section times branching fraction ($\sigma \mathcal{B}$) into four \PQb~quarks for the process \Xhy, using the asymptotic modified frequentist method (asymptotic \CLs)~\cite{Junk:1999kv, Read:2002hq, Cowan:2010js}. Figure~\ref{fig:limits} shows the expected and observed 95\% \CL upper limits as functions of \mX and \mY.
The largest excess of the observed $\sigma \mathcal{B}$, compared to the expected, is found to have a local (global) significance of 3.47 (2.44) standard deviations for the signal mass hypothesis \mX = 600\GeV and \mY = 400\GeV. The largest deficit is found to have a local\,(global) significance of $-4.62$ ($-2.65$) standard deviations for the signal mass hypothesis \mX = 600\GeV and \mY = 150\GeV. The global significance values are determined with a dedicated ``look elsewhere effect'' study \cite{Gross:2010qma,LEE2D} across the full plane of mass hypotheses used in the initial search. The observed limits are consistent with the expected ones.
Results are also shown separately for the special case of \Xhh in Fig.~\ref{fig:HHlimits}.

\begin{figure}[htb!]
\centering
{\includegraphics[width=0.49\textwidth] {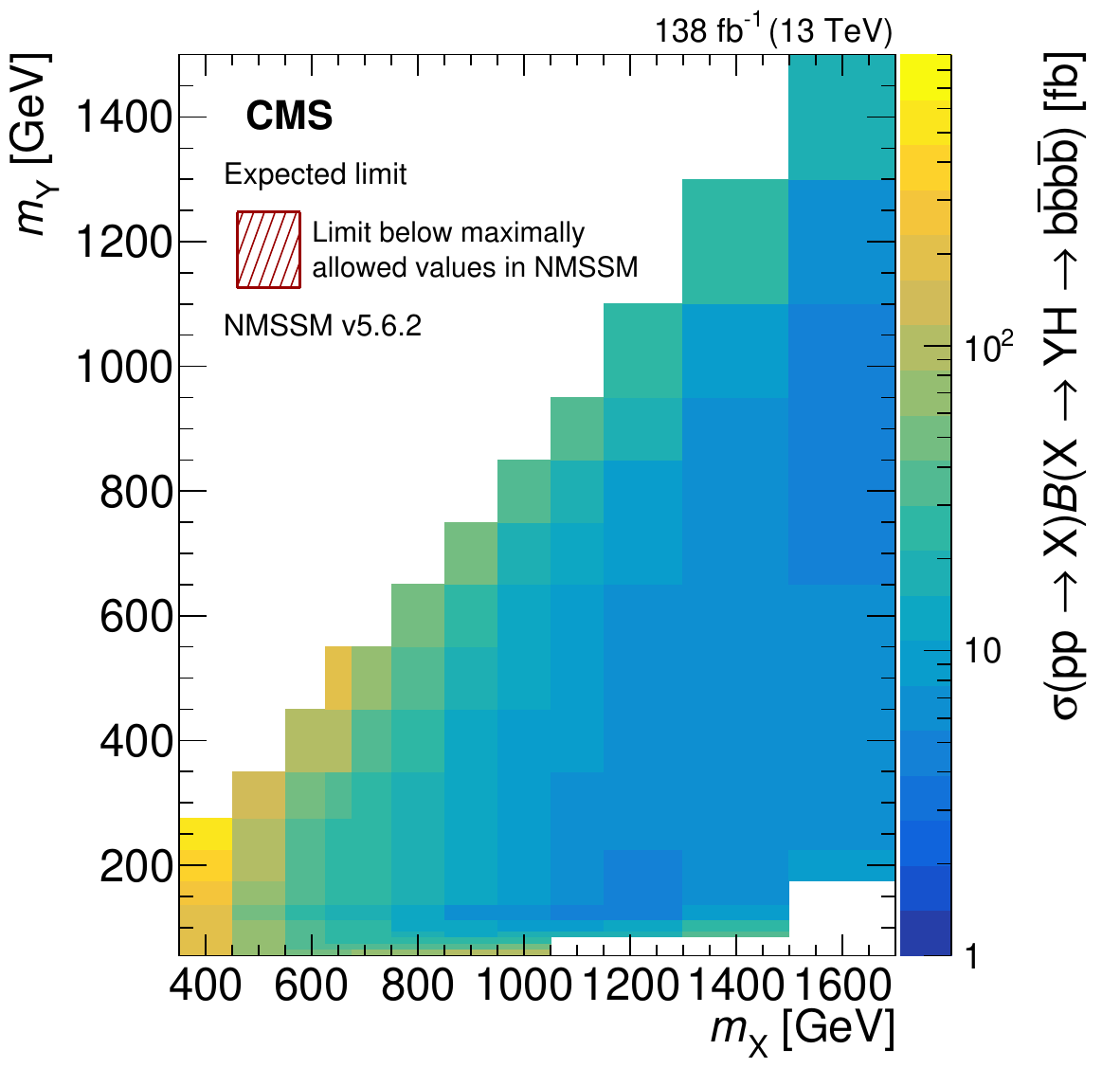}}\hspace{\fill}
{\includegraphics[width=0.49\textwidth] {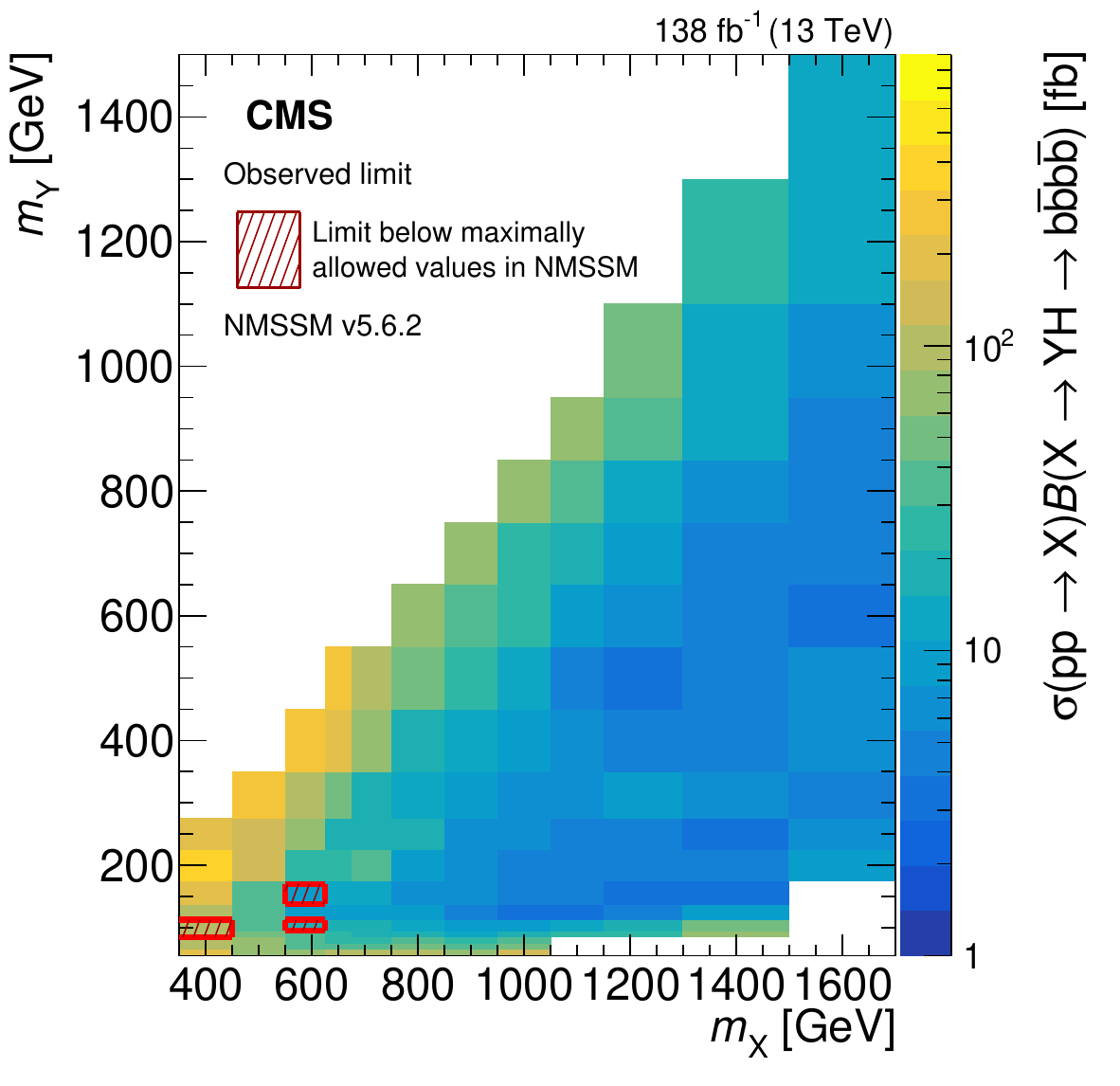}}
\caption{\label{fig:limitsOnNMSSM}
    Expected (\cmsLeft) and observed (\cmsRight) 95\% CL upper limits on $\sigma(\Pp\Pp \to \PX) \mathcal{B} (\XHY)$ shown in the (\mX,~\mY) plane.
    These limits are compared to the maximally allowed $\sigma(\Pp\Pp \to \PX) \mathcal{B} (\XHY)$ values determined with NMSSM and taking into account previous experimental constraints.
    The NMSSM limits are obtained with \textsc{NMSSMTools} 5.6.2 and appear in Ref.~\cite{Ellwanger:2022jtd}.
    A few mass hypotheses where the observed limits are more restrictive than the NMSSM limits are indicated by the red hatched areas.
}
\end{figure}

In Fig.~\ref{fig:limitsOnNMSSM}, exclusion limits are compared to the maximum $\sigma \mathcal{B}$ allowed in the NMSSM after experimental constraints from Higgs boson measurements, searches for supersymmetry, flavor physics, and dark-matter detection are accounted for. The NMSSM exclusion limits are based on \textsc{NMSSMTools} version 5.6.2 and are obtained from the scans provided in Ref.~\cite{Ellwanger:2022jtd}.

The CMS Collaboration previously reported a local (global) excess of 3.8\,(2.8) standard deviations in the $\bbbar \PGg\PGg$ channel at $\mX = 650$ and~$\mY = 90\GeV$~\cite{CMS:2023boe}. No excess is observed for this mass hypothesis in the present analysis.

Many of these results are the most stringent limits produced by CMS to date. The improvements are observed across a broad range of the mass hypotheses being investigated. For \mX mass points up to 1000\GeV, the improved expected limits are 30\% more stringent than those reported using the combination of other channels~\cite{CMS:2024phk}, on average. For \mX mass points above 1000\GeV, the improved expected limits are 84\% more stringent, on average. For some mass points, the improvements reach 77\% for $\mX \le 1000\GeV$ and 92\% for $\mX > 1000\GeV$. The \HH results are compatible with previously reported CMS limits~\cite{hh4b}. Recent ATLAS results~\cite{ATLAS:HH4b} show 50-80\% more stringent limits than the \HH results reported here.

\clearpage

\section{Summary}

This analysis presents a search for a new heavy scalar resonance (\PX) decaying into the 125\GeV standard model Higgs boson and a new scalar particle (\PY), in the \bbbb decay channel.
The search explores a range of masses from 400\GeV to 1.6\TeV for X and from 60\GeV up to 1.4\TeV for Y. A data sample corresponding to an integrated luminosity of 138\fbinv collected in proton-proton collisions at $\sqrt{s} = 13\TeV$ is used for the search. No evidence for a new signal is observed. Upper limits on the signal cross section times branching fraction are set at the 95\% confidence level. Results are interpreted in the context of the next-to-minimal supersymmetric standard model scenario and constrain the phase space of this model beyond previous experimental exclusion limits.

\begin{acknowledgments}
We congratulate our colleagues in the CERN accelerator departments for the excellent performance of the LHC and thank the technical and administrative staffs at CERN and at other CMS institutes for their contributions to the success of the CMS effort. In addition, we gratefully acknowledge the computing centres and personnel of the Worldwide LHC Computing Grid and other centres for delivering so effectively the computing infrastructure essential to our analyses. Finally, we acknowledge the enduring support for the construction and operation of the LHC, the CMS detector, and the supporting computing infrastructure provided by the following funding agencies: SC (Armenia), BMBWF and FWF (Austria); FNRS and FWO (Belgium); CNPq, CAPES, FAPERJ, FAPERGS, and FAPESP (Brazil); MES and BNSF (Bulgaria); CERN; CAS, MoST, and NSFC (China); MINCIENCIAS (Colombia); MSES and CSF (Croatia); RIF (Cyprus); SENESCYT (Ecuador); ERC PRG and PSG, TARISTU24-TK10 and MoER TK202 (Estonia); Academy of Finland, MEC, and HIP (Finland); CEA and CNRS/IN2P3 (France); SRNSF (Georgia); BMFTR, DFG, and HGF (Germany); GSRI (Greece); NKFIH (Hungary); DAE and DST (India); IPM (Iran); SFI (Ireland); INFN (Italy); MSIT and NRF (Republic of Korea); MES (Latvia); LMTLT (Lithuania); MOE and UM (Malaysia); BUAP, CINVESTAV, CONACYT, LNS, SEP, and UASLP-FAI (Mexico); MOS (Montenegro); MBIE (New Zealand); PAEC (Pakistan); MES, NSC, and NAWA (Poland); FCT (Portugal); MESTD (Serbia); MICIU/AEI and PCTI (Spain); MOSTR (Sri Lanka); Swiss Funding Agencies (Switzerland); MST (Taipei); MHESI (Thailand); TUBITAK and TENMAK (T\"{u}rkiye); NASU (Ukraine); STFC (United Kingdom); DOE and NSF (USA).

\hyphenation{Rachada-pisek} Individuals have received support from the Marie-Curie programme and the European Research Council and Horizon 2020 Grant, contract Nos.\ 675440, 724704, 752730, 758316, 765710, 824093, 101115353, 101002207, 101001205, and COST Action CA16108 (European Union); the Leventis Foundation; the Alfred P.\ Sloan Foundation; the Alexander von Humboldt Foundation; the Science Committee, project no. 22rl-037 (Armenia); the Fonds pour la Formation \`a la Recherche dans l'Industrie et dans l'Agriculture (FRIA) and Fonds voor Wetenschappelijk Onderzoek contract No. 1228724N (Belgium); the Beijing Municipal Science \& Technology Commission, No. Z191100007219010, the Fundamental Research Funds for the Central Universities, the Ministry of Science and Technology of China under Grant No. 2023YFA1605804, the Natural Science Foundation of China under Grant No. 12535004, and USTC Research Funds of the Double First-Class Initiative No.\ YD2030002017 (China); the Ministry of Education, Youth and Sports (MEYS) of the Czech Republic; the Shota Rustaveli National Science Foundation, grant FR-22-985 (Georgia); the Deutsche Forschungsgemeinschaft (DFG), among others, under Germany's Excellence Strategy -- EXC 2121 ``Quantum Universe" -- 390833306, and under project number 400140256 - GRK2497; the Hellenic Foundation for Research and Innovation (HFRI), Project Number 2288 (Greece); the Hungarian Academy of Sciences, the New National Excellence Program - \'UNKP, the NKFIH research grants K 131991, K 133046, K 138136, K 143460, K 143477, K 146913, K 146914, K 147048, 2020-2.2.1-ED-2021-00181, TKP2021-NKTA-64, and 2025-1.1.5-NEMZ\_KI-2025-00004 (Hungary); the Council of Science and Industrial Research, India; ICSC -- National Research Centre for High Performance Computing, Big Data and Quantum Computing, FAIR -- Future Artificial Intelligence Research, and CUP I53D23001070006 (Mission 4 Component 1), funded by the NextGenerationEU program, the Italian Ministry of University and Research (MUR) under Bando PRIN 2022 -- CUP I53C24002390006, PRIN PRIMULA 2022RBYK7T (Italy); the Latvian Council of Science; the Ministry of Education and Science, project no. 2022/WK/14, and the National Science Center, contracts Opus 2021/41/B/ST2/01369, 2021/43/B/ST2/01552, 2023/49/B/ST2/03273, and the NAWA contract BPN/PPO/2021/1/00011 (Poland); the Funda\c{c}\~ao para a Ci\^encia e a Tecnologia (Portugal); the National Priorities Research Program by Qatar National Research Fund; MICIU/AEI/10.13039/501100011033, ERDF/EU, ``European Union NextGenerationEU/PRTR", projects PID2022-142604OB-C21, PID2022-139519OB-C21, PID2023-147706NB-I00, PID2023-148896NB-I00, PID2023-146983NB-I00, PID2023-147115NB-I00, PID2023-148418NB-C41, PID2023-148418NB-C42, PID2023-148418NB-C43, PID2023-148418NB-C44, PID2024-158190NB-C22, RYC2021-033305-I, RYC2024-048719-I, CNS2023-144781, CNS2024-154769 and Plan de Ciencia, Tecnolog{\'i}a e Innovaci{\'o}n de Asturias, Spain; the Chulalongkorn Academic into Its 2nd Century Project Advancement Project, the National Science, Research and Innovation Fund program IND\_FF\_68\_369\_2300\_097, and the Program Management Unit for Human Resources \& Institutional Development, Research and Innovation, grant B39G680009 (Thailand); the Eric \& Wendy Schmidt Fund for Strategic Innovation through the CERN Next Generation Triggers project under grant agreement number SIF-2023-004; the Kavli Foundation; the Nvidia Corporation; the SuperMicro Corporation; the Welch Foundation, contract C-1845; and the Weston Havens Foundation (USA).
\end{acknowledgments}\section*{Data availability} Release and preservation of data used by the CMS Collaboration as the basis for publications is guided by the  \href{https://doi.org/10.7483/OPENDATA.CMS.1BNU.8V1W}{CMS data preservation, re-use and open access policy}.

\bibliography{auto_generated}

\cleardoublepage \appendix\section{The CMS Collaboration \label{app:collab}}\begin{sloppypar}\hyphenpenalty=5000\widowpenalty=500\clubpenalty=5000\cmsinstitute{Yerevan Physics Institute, Yerevan, Armenia}
{\tolerance=6000
A.~Hayrapetyan, V.~Makarenko\cmsorcid{0000-0002-8406-8605}, A.~Tumasyan\cmsAuthorMark{1}\cmsorcid{0009-0000-0684-6742}
\par}
\cmsinstitute{Institut f\"{u}r Hochenergiephysik, Vienna, Austria}
{\tolerance=6000
W.~Adam\cmsorcid{0000-0001-9099-4341}, L.~Benato\cmsorcid{0000-0001-5135-7489}, T.~Bergauer\cmsorcid{0000-0002-5786-0293}, M.~Dragicevic\cmsorcid{0000-0003-1967-6783}, C.~Giordano\cmsorcid{0000-0001-6317-2481}, P.S.~Hussain\cmsorcid{0000-0002-4825-5278}, M.~Jeitler\cmsAuthorMark{2}\cmsorcid{0000-0002-5141-9560}, N.~Krammer\cmsorcid{0000-0002-0548-0985}, A.~Li\cmsorcid{0000-0002-4547-116X}, D.~Liko\cmsorcid{0000-0002-3380-473X}, M.~Matthewman, J.~Schieck\cmsAuthorMark{2}\cmsorcid{0000-0002-1058-8093}, R.~Sch\"{o}fbeck\cmsAuthorMark{2}\cmsorcid{0000-0002-2332-8784}, M.~Shooshtari\cmsorcid{0009-0004-8882-4887}, M.~Sonawane\cmsorcid{0000-0003-0510-7010}, W.~Waltenberger\cmsorcid{0000-0002-6215-7228}, C.-E.~Wulz\cmsAuthorMark{2}\cmsorcid{0000-0001-9226-5812}
% ...
\par}
\vskip\cmsinstskip
\dag:~Deceased\\
$^{1}$Also at Yerevan State University, Yerevan, Armenia\\
$^{2}$Also at TU Wien, Vienna, Austria\\
% ...
\end{sloppypar}
\end{document}